\documentclass[aps,prd,amssymb,eqsecnum,showpacs]{revtex4}
\usepackage{graphicx,psfrag}

\begin{document}

\title{Testing numerical evolution with the shifted gauge wave}

\author{Maria C. Babiuc${}^{1}$ ,
	B\'{e}la Szil\'{a}gyi${}^{2}$ ,
	Jeffrey Winicour${}^{1,2}$
       }
\affiliation{
${}^{1}$ Department of Physics and Astronomy \\
         University of Pittsburgh, Pittsburgh, PA 15260, USA \\
${}^{2}$ Max-Planck-Institut f\" ur
         Gravitationsphysik, Albert-Einstein-Institut, \\
	 14476 Golm, Germany
	 }
\today

\begin{abstract}

Computational methods are essential to provide waveforms from coalescing black
holes, which are expected to produce strong signals for the gravitational wave
observatories being developed. Although partial simulations of the coalescence
have been reported, scientifically useful waveforms have so far not been
delivered. The goal of the AppleswithApples (AwA) Alliance is to design,
coordinate and document standardized code tests for comparing numerical
relativity codes. The first round of AwA tests have now being completed and the
results are being analyzed. These initial tests are based upon periodic
boundary conditions designed to isolate performance of the main evolution code.
Here we describe and carry out an additional test with periodic boundary
conditions which deals with an essential feature of the black hole excision
problem, namely a non-vanishing shift. The test is a shifted version of the
existing AwA gauge wave test. We show how a shift introduces an exponentially
growing instability which  violates the constraints of a standard harmonic
formulation of Einstein's equations. We analyze the Cauchy problem in a
harmonic gauge and discuss particular options for suppressing instabilities in
the  gauge wave  tests. We implement these techniques in a finite difference
evolution algorithm and present test results. Although our application here is
limited to a model problem, the techniques should benefit the simulation of
black holes  using harmonic evolution codes.

\end{abstract}


\maketitle

\section{Introduction}

Computational methods are essential to provide the waveform from the
coalescence of black holes, which is expected to produce a strong signal
for the gravitational wave observatories being developed.  The importance
of the binary black hole problem to the success of LIGO and LISA has led
to major computational efforts, most notably the Binary Black Hole Grand
Challenge. Although this Grand Challenge had intermediate
successes~\cite{sci,gce,mbh}, scientifically useful waveforms were not
delivered. At present, this remains a problem beyond the reach of any
existing code.

A recent study~\cite{kendpost} of large scale scientific code projects at the
Livermore, Los Alamos and Sandia National Laboratories, funded under the U.S.
Department of Energy's Accelerated Strategic Computing Initiative (ASCI),
identified three {\em necessary} elements for success: {\em verification,
validation and quality management}. In the absence of any of those three
requirements, the report concluded that the results would have little
scientific impact because of the impossibility to judge code reliability.
Although the ASCI projects involved highly experienced and qualified teams at
laboratories with ample resources, only one third of the projects succeeded as
planned, another third succeeded later than planned and the remaining were
eventually abandoned. The failed projects had overly ambitious schedules and
goals and lacked a conservative methodology that minimized risk. It was
expected that the failure rate would have been much higher for simulation
projects without the experience and resources of the ASCI teams.

The lessons learned from the ASCI and other studies have been reviewed by Post
and Votta~\cite{postvot}. Their experience with the peer review process in
computational science was that it is not as effective as in experiment or
theory, because the many possible sources of hidden defects make the referee
rely too heavily on plausibility checks as opposed to independent reproduction
of results. Some of their observations, which are presented below in quotes,
are very pertinent to numerical relativity: ``New methods of verifying and
validating complex codes are mandatory if computational science is to fulfill
its promise for science and society''. The {\em validation} of a code implies
that the predictions of the code are in accord with observed phenomena. For the
present status of the binary black hole problem, in the absence of any
empirical observations, the burden falls completely on {\em verification}. 

Post and Votta list five verification techniques, each one having limited effectiveness by itself:

\begin{enumerate}

\item ``Comparing code results with an exact answer''.

\item ``Establishing that the convergence rate of the truncation error
with changing grid spacing is consistent with expectations''.

\item ``Comparing calculated with expected results for a problem especially
manufactured to test the code''.

\item ``Monitoring conserved quantities and parameters, preservation
of symmetry properties and other easily predictable outcomes''.

\item ``Benchmarking -- that is, comparing results from those with
existing codes that can calculate similar problems''.

\end{enumerate}

The importance of the first four techniques has been recognized by most
numerical relativity groups and their implementation in practice has improved
the integrity of the field. Individual groups cannot easily carry out the fifth
technique independently. This was the motivation behind formation of the
AppleswithApples (AwA) Alliance, whose goal  is to design, coordinate and
document standardized code tests for comparing numerical relativity codes. A
pivotal step has been the documentation of a first round of AwA tests based
upon periodic boundary condition (equivalent to a 3-torus with no
boundary)~\cite{mex1}, designed to isolate performance of the main evolution
code. The choice of initial AwA tests was biased by considerations of
expediency arising from the state of the field at the time. The cross
fertilization between computational mathematics and numerical relativity was
just entering a productive stage. Only a few groups had based their codes upon
well-posed symmetric or strongly hyperbolic formulations of Einstein's
equations and fewer groups had an understanding of how to treat boundary
conditions. Detailed specifications of a second round of tests, involving
boundaries  (essential for the black hole excision problem), have now been
proposed~\cite{awa}. 

Post and Votta emphasized that

\begin{itemize}

\item ``Verification and validation establish the credibility of code
predictions. Therefore it's very important to have a written record of
verification and validation results.''

\item ``Without such programs, computational science will never be
credible.''

\end{itemize}

The first round of AwA tests have now been completed using codes based upon
most of the prominent numerical relativity formalisms. Results carried out by
the participating groups can be viewed in the Alliance CVS depository.
Instructions for accessing the data in the CVS are available at the Alliance
website~\cite{awa}. A second paper is in preparation which discusses these
first round test results with respect to code performance and improvements in
the test specifications~\cite{mex2}.

Here we describe and carry out an additional test based upon periodic boundary
conditions which deals with an essential feature of the black hole excision
problem, namely a non-vanishing shift. The test is a shifted version of the
existing AwA gauge wave test. We detail the test specifications in
Sec.~\ref{sec:sgw}. 

In Sec.~\ref{sec:gcv}, we describe some instabilities associated with the gauge
wave and shifted gauge wave tests. A discussion of the Cauchy problem for
general relativity depends upon a choice of gauge conditions which reduce
Einstein's equations to hyperbolic form. See~\cite{friedrend} for a recent
review. The simplest reduction is in terms of harmonic
coordinates~\cite{dedonder,Fock}, for which well-posedness of the Cauchy
problem was first established~\cite{Choquet}. Previous work~\cite{bab} has
revealed that the gauge wave without shift has a constraint preserving
instability in harmonic coordinates, i.e the gauge wave metric has
exponentially growing perturbations which satisfy the harmonic conditions and
Einstein's equations. We show here how a shift introduces a new type of
exponentially growing instability in the standard harmonic reduction of
Einstein's equations. 

In Sec.~\ref{sec:cauchy}, we summarize various options for constructing a
hyperbolic evolution algorithm based upon the harmonic formulation. Our
discussion centers around the standard form of the harmonic formulation adopted
in most recent analytic treatments. Although the well-posedness of the Cauchy
problem guarantees the existence and uniqueness of a solution with continuous
dependence on the initial data, this is only a necessary condition for
computational success and is not sufficient. Simulations have shown that
instabilities in the gauge wave metrics can quickly introduce unacceptable
error even with a convergent code~\cite{bab}. In the case of the gauge wave
without shift, these instabilities can be suppressed by implementing discrete
conservation laws obeyed by the principle part of the evolution
system~\cite{bab}. However, in the case of the shifted gauge wave this
technique is not effective by itself because the instability is excited by
nonlinear terms. In Sec.~\ref{sec:cauchy}, we discuss options at the analytic
level for suppressing this instability in the shifted gauge wave test. One
option is to adjust the nonlinear terms in the evolution system by adding terms
which vanish modulo the harmonic constraints. Another option is the inclusion
of harmonic gauge forcing terms,  which in principle allow the simulation of
any nonsingular spacetime region. Gauge forcing terms not only allow the
flexibility to ``steer'' around coordinate pathologies that might arise but
they also allow the adaptability to carry out standardized tests in any
specified gauge. 

Section \ref{sec:imp} describes the implementation of the harmonic formulation
as a finite difference evolution code. We base our approach on a
code~\cite{harl}, the Abigel code which was developed to implement a
well-posed, constraint preserving version of the harmonic initial-boundary
value problem (IBVP). Here we confine our attention to the Cauchy problem. In
future work we will extend our test results to the IBVP. Since the preliminary
testing of the Abigel code, considerable improvement has  been made in the
underlying numerical techniques. The study of a model nonlinear scalar 
wave~\cite{excis} shows that semi-discrete conservation laws and summation by
parts can be used to formulate stable algorithms. In Section \ref{sec:imp}, we
generalize these techniques to the reduced harmonic Einstein equations. These
techniques also have potential application to the $Z4$ system~\cite{z4} which
generalizes the harmonic formulation.

In Sec.~\ref{sec:tests}, we calibrate the stability,  convergence and
performance of the code using the gauge wave and shifted gauge wave metrics. We
show how numerical noise excites  instabilities that can be cured by a
combination of discrete conservation laws and constraint adjustments.

\section{The shifted gauge wave test}
\label{sec:sgw}

The standard AwA gauge wave test is based upon
the flat metric 
\begin{equation}
  ds^2=(1-H)(- dt^2 +dx^2)+dy^2+dz^2,
\label{eq:gw}
\end{equation}
where
\begin{equation}
  \label{eq:flatgaugewaveHfn}
  H = H(x-t)= A \sin \left( \frac{2 \pi (x - t)}{d} \right)
\label{eq:H}
\end{equation}
is a sinusoidal wave of amplitude $A$ propagating
along the $x$-axis. In order to test 2-dimensional features, 
the coordinates are rotated according to 
\begin{equation}
   x = \frac{1}{\sqrt{2}}(x^\prime - y^\prime), \qquad
    y = \frac{1}{\sqrt{2}}(x^\prime + y^\prime) \, 
 \label{eq:flatgaugewave1to2d}
\end{equation}
which produces a gauge wave
propagating along the diagonal with dependence
\begin{equation}
\sin \left( \frac{2 \pi (x' - y' - t' \sqrt{2})}{d'} \right), 
\quad \textrm{where} \quad d' = d \sqrt{2} \, .
\end{equation}
Adjusting $d$ or $d'$ to the size of the evolution domain gives periodicity in
the $x$ and $y$ directions.

The shifted gauge wave is obtained from the Minkowski metric
$ds^2 =- d\hat t^2+d\hat x^2+d\hat y^2+d\hat z^2$ by the coordinate
transformation 
\begin{equation}
  \label{eq:GaugeWave1}
  \begin{array}[c]{r c l}
   \hat t&=& t - \frac {Ad}{4\pi}\cos \left( \frac{2 \pi (x - t)}{d} \right), 
      \\
   \hat x&=& x - \frac {Ad}{4\pi}\cos \left( \frac{2 \pi (x - t)}{d} \right), 
      \\
     \hat y&=& y,    \\ 
     \hat z&=& z
  \end{array}
\end{equation}
where $d$ is the size of the evolution domain. This leads to the 
4-metric of Kerr-Schild form
\begin{equation}
  \label{eq:flatgaugewave4metric}
  ds^2=- dt^2 +dx^2+dy^2+dz^2 +H k_\alpha k_\beta dx^\alpha dx^\beta
\label{eq:sgw}
\end{equation}
where
\begin{equation}
      k_\alpha=\partial_\alpha (x-t) =(-1,1,0,0)
\end{equation}
and $H$ is again given by (\ref{eq:H}). This metric describes a shifted gauge
wave of amplitude $A$ propagating along the $x$-axis.  As above,  the
coordinate transformation (\ref{eq:flatgaugewave1to2d}) rotates the propagation
direction to the diagonal. 

The shifted gauge wave test is run in both axis-aligned 1D form and diagonal 2D
form. We run the test with amplitude $A=0.5$. We have found that smaller
amplitudes are not as efficient for revealing problems. Larger amplitudes can
trigger gauge pathologies, e.g $g^{tt}\ge 0$ (breakdown of the spacelike nature
of the Cauchy hypersurfaces), more quickly and complicate code comparisons.

We specify the wavelength $d=1$ in the 1D simulations and $d'=\sqrt{2}$ in the
2D simulations. We find that at least 50 grid points lead to reasonable
simulations for more than 10 crossing times and therefore make the following
choices for the computational grid:
\begin{itemize}
\item Simulation domain:
\begin{center}
\begin{tabular}{rllll}
  1D:& $\quad  x \in [-0.5, +0.5],$ & $\quad y = 0,$ &$ \quad z = 0,$
  & $ \quad d=1$ \\
  2D: & $\quad x \in [-0.5, +0.5], $ & 
    $\quad y \in [-0.5, +0.5], $ & $\quad z=0,$ & $\quad d'=\sqrt{2}$
\end{tabular}
\end{center}
\item Grid: $x_n = -0.5 + n dx, \quad n=0,1\ldots 50\rho,
  \quad dx=dy=dz=1/(50\rho), \quad \rho = (1,2,4)$
\item Time step: $dt = dx/4 = 0.005 / \rho$
\end{itemize}
The 1D evolution is carried out for $T=1000$ crossing times,
i.e.~$2\times10^5\rho$ time steps (or until the code crashes). The 2D diagonal
runs are carried out for $T=100$.

Useful output data are the profiles along the $x$-axis through the center of
the grid $(y=z=0)$ of $g_{tt}$, $g_{xt}$, and $g_{xx}$, the $\ell_2$ and
$\ell_\infty$ norms of the error and the Hamiltonian and momentum constraints,
or any other constraints which arise in a particular formulation. It is also
important to calculate the convergence factor of the numerical error. 

\section{Instability of the shifted gauge wave}
\label{sec:gcv}

Both the gauge wave metric (\ref{eq:gw}) and shifted gauge wave metric
(\ref{eq:sgw}) are  flat vacuum solutions of the Einstein equation in
harmonic coordinates 
$\Box x^\mu =-\Gamma^\mu=0$, where
\begin{equation}
     \Gamma^\mu = g^{\alpha\beta}\Gamma^\mu_{\alpha\beta}= 
              -\frac{1}{\sqrt {-g}}\partial_\alpha\gamma^{\alpha\mu}
\end{equation} 
and $\gamma^{\mu\nu}=\sqrt{-g}g^{\mu\nu}$.

Simulation of the AwA gauge wave without shift (\ref{eq:gw}) is complicated by
the related metric~\cite{bab}
\begin{equation}
  ds_\lambda^2=e^{\lambda t}(1-H)(- dt^2 +dx^2)+dy^2+dz^2
\label{eq:lgw}
\end{equation}
which, for any value of $\lambda$, is a flat metric which
obeys the harmonic coordinate conditions and thus represents a
harmonic gauge instability of Minkowski space with periodic boundary conditions.

The shifted gauge wave (\ref{eq:sgw}) has an analogous instability
\begin{equation}
  ds_\lambda^2=e^{\lambda t}(- dt^2 +dx^2)+dy^2+dz^2
           +H k_\alpha k_\beta dx^\alpha dx^\beta,
\label{eq:lsgw}
\end{equation}
which is again a vacuum metric. However, the metric (\ref{eq:lsgw}) is
not in harmonic coordinates and has a harmonic driving term~\cite{Friedrich}
$\Gamma^\alpha =\hat \Gamma^\alpha$, where 
\begin{equation}
     \hat \Gamma^\alpha = -\lambda H k^\alpha.
\label{eq:egauge}   
\end{equation}
Thus this particular instability would only be excited in a
gauge satisfying (\ref{eq:egauge}). In the standard $3+1$ description
$x^\mu=(t,x^i)$, this determines the propagation 
of the lapse $\alpha=1/\sqrt{-g^{tt}}$ and shift
$\beta^i=-g^{it}/g^{tt}$ according to
\begin{equation}
  \frac{1}{\alpha\sqrt{h}}\bigg (\partial_t \bigg (
        \frac {\sqrt{h}}{\alpha} \bigg ) 
	  -\partial_j \bigg ( \frac {\sqrt{h}\beta^j}{\alpha} \bigg ) 
       \bigg ) =\hat \Gamma^t
\label{eq:lapse}
\end{equation}
and
\begin{equation}
   -\frac{1}{\alpha\sqrt{h}}\bigg (
           \partial_t \bigg ( \frac {\sqrt{h}\beta^i}{\alpha} \bigg )
	   +\partial_j \bigg (\alpha\sqrt{h}(h^{ij}
	      -\frac{\beta^i \beta^j}{\alpha^2})
                              \bigg )  \bigg )= \hat \Gamma^i,
\label{eq:shift}
\end{equation}
where $h_{ij}$ is the 3-metric and $h=\det(h_{ij})$. 

The exponentially growing metrics (\ref{eq:lgw}) and (\ref{eq:lsgw}) both
satisfy the Einstein equations. However, the shifted gauge wave also has a
different type of instability when the evolution system is taken to be the
standard harmonic reduction of  the Einstein tensor~\cite{friedrend,Wald}
\begin{equation}
     E^{\mu\nu}:= G^{\mu\nu} -\nabla^{(\mu}\Gamma^{\nu)} 
                 +\frac{1}{2}g^{\mu\nu}\nabla_\alpha \Gamma^\alpha,
		 \label{eq:e}
\end{equation}
which leads to the hyperbolic evolution equation
\begin{equation}
     E^{\mu\nu}=0 .
 \label{eq:eeq}
\end{equation}
(Here $\Gamma^\nu$ is treated formally as a vector in constructing the
``covariant'' derivative $\nabla^{\mu}\Gamma^{\nu}$).  In order to see the
origin of this instability first consider the spatially homogeneous case with
amplitude $A=0$, for which the shifted gauge wave metric reduces to the
Minkowski metric $\eta_{\mu\nu}$. Nonlinear perturbations of this  metric of
the Kerr-Schild form  
\begin{equation}
     g_{\mu\nu}=\eta_{\mu\nu} +F(t) k_\mu k_\nu
\label{eq:Fks}
\end{equation}
(where $k_\alpha =\partial_\alpha(x-t)$ as before) satisfy
the reduced harmonic equations (\ref{eq:eeq}) provided
\begin{equation}
  E^{\mu\nu} 
       =\frac{1}{2}\bigg( (1+F) F_{,tt} -F_{,t} F_{,t} \bigg)k^\mu k^\nu =0.
  \label{eq:expkst}
\end{equation}
This has the exponential solution
\begin{equation}
       F=e^{\lambda t}-1.
\label{eq:F}
\end{equation}
The resulting metric
\begin{equation}
     g_{\mu\nu}=\eta_{\mu\nu} +(e^{\lambda t}-1) k_\mu k_\nu
\label{eq:expt}
\end{equation}
solves the reduced equations but unlike (\ref{eq:lsgw}) it does not solve the
Einstein equations because the harmonic condition is not satisfied. Instead
\begin{equation}
     \Gamma^\mu = \lambda e^{\lambda t}k^\mu.
\label{eq:gv}
\end{equation}
From (\ref{eq:e}), this implies that $G^{t\alpha}=0$, i.e. that
the Hamiltonian and momentum constraints are satisfied, but that
the Einstein tensor has the non-vanishing components
\begin{equation}
    G^{yy}= G^{zz}= -\frac{1}{2}\lambda^2 e^{\lambda t}.
\end{equation}

Whether the unstable mode (\ref{eq:expt}) is excited by numerical error
depends upon the evolution system. A system which enforces $G^{yy}= G^{zz}=0$
would not excite this mode. In the case of harmonic evolution, the Einstein
equations are satisfied only indirectly through the satisfaction of the
harmonic conditions. These conditions, $\Gamma^\mu=0$, are the constraints of
the harmonic evolution system. Because $\Gamma^\mu$ is not an evolution
variable, error of the form (\ref{eq:gv}) can be expected to
excite the instability.
 
The shifted gauge wave (\ref{eq:sgw}) has an exponentially growing instability
analogous to (\ref{eq:expt}) that satisfies the reduced harmonic equations
but violates the harmonic constraints. The
unstable perturbation may be constructed analytically by applying the
transformation (\ref{eq:GaugeWave1}) to the exponential solution
(\ref{eq:expt}) of the reduced harmonic equations. Because this transformation
has the property $\hat x -\hat t =x-t$, it is simple to verify that
$\Gamma^\mu$ transforms as a vector. As a result, the reduced equations
(\ref{eq:e}) remain satisfied (since $G^{\mu\nu}$ is a tensor). The resulting
metric is
\begin{equation}
  \label{eq:lsgaugewave4metric}
  ds_\lambda^2=- dt^2 +dx^2+dy^2+dz^2 +
        \bigg(H-1 +e^{\lambda \hat t}\bigg )
            k_\alpha k_\beta dx^\alpha dx^\beta ,
\label{eq:instab}
\end{equation}
where now
\begin{equation}
\hat t= t - \frac {Ad}{4\pi}\cos \left( \frac{2 \pi (x - t)}{d} \right) .
\end{equation}
The resulting harmonic constraint violation is given by
\begin{equation}
     \Gamma^\mu =\lambda e^{\lambda\hat t} k^\mu.
\end{equation}
The simulation of the shifted gauge wave by a harmonic evolution algorithm
based upon the standard reduction (\ref{eq:e}) excites this instability, as
exhibited by the results in \ref{sec:tests}. However, the instability can be
controlled by modifying the standard harmonic system, as discussed in the next
section.    

\section{The harmonic Cauchy problem}
\label{sec:cauchy}

The standard harmonic reduction of the Einstein equations is given by (\ref{eq:e}), with $\Gamma^\mu$ set to zero, which leads
to a hyperbolic system (\ref{eq:eeq}) which can be cast into the flux conservative form
\begin{eqnarray}
    2\sqrt{-g} E^{\mu\nu}&=&\partial_\alpha (g^{\alpha\beta}
           \partial_\beta \gamma^{\mu\nu})  
   -2\sqrt{-g}g^{\alpha\rho}g^{\beta\sigma}\Gamma^\mu_{\alpha\beta}
                   \Gamma^{\nu}_{\rho\sigma}
	-\sqrt{-g}(\partial_\alpha g^{\alpha\beta})\partial_\beta g^{\mu\nu}
   +\frac{1}{\sqrt{-g}}g^{\alpha\beta}(\partial_\beta g)\partial_\alpha g^{\mu\nu} 
             \nonumber \\
      &+&\frac{1}{2}g^{\mu\nu}\bigg (\frac{1}{2g\sqrt{-g}}g^{\alpha\beta}
                 (\partial_\alpha g)\partial_\beta g
        +\sqrt{-g}\Gamma^\rho_{\alpha\beta}\partial_\rho g^{\alpha\beta}
	+\frac{1}{\sqrt{-g}}(\partial_\beta g)\partial_\alpha g^{\alpha\beta}
	   \bigg )=0.
\label{eq:efc}
\end{eqnarray}
The harmonic condition $\Gamma^\mu=0$ comprise the constraints which are
sufficient to establish that the Einstein tensor vanishes. When the reduced
harmonic equations (\ref{eq:efc}) are satisfied, the Bianchi identities imply 
\begin{equation}
     \nabla^\alpha \nabla_\alpha \Gamma^\mu +R^\mu_\nu \Gamma^\nu =0,
     \label{eq:constr} 
\end{equation}
where the Ricci tensor satisfies
\begin{equation}
     R^{\mu\nu} = \nabla^{(\mu}\Gamma^{\nu)}.
     \label{eq:redricci}
\end{equation}
These equations provide the key result that the harmonic conditions propagate
in time if the reduced equations are satisfied.  The historic proof of
well-posedness of the initial value problem for Einstein's
equations~\cite{Choquet} follows from the  hyperbolicity of (\ref{eq:e}) and
(\ref{eq:constr}). Here hyperbolicity can be defined either in terms of the
second differential order systems (\ref{eq:efc}) and (\ref{eq:constr}), as
in~\cite{Choquet}, or by reducing (\ref{eq:efc}) and (\ref{eq:constr}) to first
order symmetric hyperbolic systems~\cite{fisher}.

The principal part of  $E^{\mu\nu}$, i.e.
\begin{equation}
     \partial_\alpha (g^{\alpha\beta} \partial_\beta \gamma^{\mu\nu}),
\label{eq:pp}
\end{equation}      
has been chosen so that when the remaining nonlinear terms in $E^{\mu\nu}$
vanish (or can be neglected) the associated conservation laws suppress
excitation of the vacuum solution (\ref{eq:lgw}). These conservation laws, when
implemented in the discretized system, successfully suppress the instability
(\ref{eq:lgw}) in the   simulation of the gauge wave without shift.
See~\cite{bab} for details. The discussion in~\cite{bab} suggests that other 
flux conservative forms of the principle part whose terms have an analogous 
conformal weight, such as 
\begin{equation}
     \partial_\alpha (g^{\alpha\beta} \partial_\beta g_{\mu\nu}),
\end{equation} 
would be equally effective at suppressing this unstable mode. 

On the other hand, we have found that these conservation laws associated with
the principle part (\ref{eq:pp}) are not effective in suppressing the
instability (\ref{eq:instab}) in the simulation of the shifted gauge wave.  In
this case, the instability is excited by the first derivative terms terms in
(\ref{eq:efc}) which act as a nonlinear source for the principle part
(\ref{eq:pp}). This instability must be handled by a different technique.
There are two straightforward generalizations of the standard harmonic
treatment which leave the principle part of (\ref{eq:efc}) unchanged, so that
the well-posedness of the Cauchy problem remains intact, but modify the
nonlinear terms: (i) the introduction of harmonic gauge source functions or
(ii) constraint adjustment of the equations.

\subsection{Harmonic gauge source functions}
 
Harmonic gauge source functions $\hat \Gamma^{\mu}(x^\alpha, g_{\alpha\beta})$,
which are explicit functions of the coordinates and the metric,
are introduced by replacing the harmonic conditions by
$\Gamma^{\mu}=\hat \Gamma^{\mu}$~\cite{Friedrich}.
The reduced equations then become 
\begin{equation}
   \hat E^{\mu\nu}=E^{\mu\nu}+\nabla^{(\mu}\hat \Gamma^{\nu)} 
                 -\frac{1}{2}g^{\mu\nu}\nabla_\alpha \hat \Gamma^\alpha
		 \label{eq:ehat} =0.
\end{equation} 
The generalized harmonic conditions 
\begin{equation}
   {\cal C}^\mu :=\Gamma^\mu -\hat \Gamma^\mu =0,
\label{eq:Gconstr}
\end{equation}
are sufficient to ensure that Einstein's equations are satisfied. For this
reason we refer to ${\cal C}^\mu$ as the constraints of the generalized
harmonic formulation. They are related to the standard Hamiltonian and momentum
constraints.

The Bianchi identities imply
\begin{equation}
     \nabla^\alpha \nabla_\alpha {\cal C}^\mu +R^\mu_\alpha {\cal C}^\alpha =0.
     \label{eq:hatconstr}
\end{equation}
Because the hyperbolicity of (\ref{eq:ehat}) and ({\ref{eq:hatconstr}) is
unaffected by such gauge source functions, the Cauchy problem remains
well-posed. The uniqueness of solutions to (\ref{eq:hatconstr}) thus ensures
that the harmonic constraints ${\cal C}^\mu=0$ are satisfied in the domain of
dependence of the initial Cauchy hypersurface ${\cal S}$ provided the initial
data satisfy ${\cal C}^\mu=\partial_t {\cal C}^\mu=0$. It is straightforward to
verify that Cauchy data on ${\cal S}$ which satisfy the Hamiltonian and
momentum constraints $G^t_\mu=0$ and the initial condition ${\cal C}^\mu=0$
also satisfy $\partial_t {\cal C}^\mu =0$ on ${\cal S}$ by virtue of the
generalized reduced harmonic equations (\ref{eq:ehat}). 

The harmonic constraints (\ref{eq:Gconstr}) also imply equations
(\ref{eq:lapse}) and (\ref{eq:shift}) governing the time derivatives of the
lapse and shift. As a result, in addition to the standard Cauchy data, i.e. the
intrinsic metric and extrinsic curvature of ${\cal S}$ subject to the
Hamiltonian and momentum constraints, the only other free initial data are the
initial choices of lapse and shift.

Thus given Cauchy data $\gamma^{\mu\nu}$ and $\partial_t \gamma^{\mu\nu}$ that
satisfies the Hamiltonian and momentum constraints in the gauge
$\Gamma^\mu=0$, Cauchy data  $\hat \gamma^{\mu\nu}$ and $\partial_t \hat
\gamma^{\mu\nu}$ that satisfies the constraints in the gauge $\Gamma^\mu=\hat
\Gamma^\mu$ can be obtained by setting $\hat \gamma^{\mu\nu}=\gamma^{\mu\nu}$ and
$\partial_t \hat g^{ij}=\partial_t g^{ij}$ and then determining $\partial_t
\hat \gamma^{t\alpha}$ from
\begin{equation}
   \frac{1}{\sqrt{-\hat g}} \partial_\nu \hat \gamma^{\mu\nu} 
    = -\hat \Gamma^\mu.
\label{eq:hat}
\end{equation}
Finally, $\partial_t \hat \gamma^{ij}$ is determined from 
\begin{equation}
     \frac{1}{\sqrt{-\hat g}}\partial_t \hat \gamma^{ij}
                =  \partial_t \hat g^{ij} 
           + \frac{1}{2\hat g}\hat g^{ij}\partial_t \hat g.
\end{equation}  
Here the calculation of $\partial_t \hat g$ can be expedited
using the identity $g^{tt} g =h$, where $h=1/\det(g^{ij})$ is the determinant
of the 3-metric $h_{ij}$. Combined with the $t$-component of (\ref{eq:hat}),
this yields
\begin{equation}
       \frac {\partial_t \hat g}{2\hat g} = \frac{\partial_t h}{h}
                 +\frac{1}{g^{tt}}(\hat \Gamma^t+
             \frac{1}{\sqrt{-g}}\partial_i \gamma^{ti}).
\end{equation}

In order to examine whether a gauge source function can be effective in
controlling the instability of the shifted gauge wave, we consider the simpler
problem of spatially homogeneous Kerr-Schild metrics (\ref{eq:Fks}).
First consider the choice 
\begin{equation}
  \hat \Gamma^\mu = c t^\mu
\end{equation}
where $t^\mu\partial_\mu = \partial_t$ is the evolution vector. The
resulting modification to (\ref{eq:expkst}) is 
\begin{equation}
  \hat E^{\mu\nu} 
       =\frac{1}{2}\bigg ( (1+F) F_{,tt} -F_{,t} F_{,t} + cF_{,t} \bigg)
                       k^\mu k^\nu =0.
\end{equation}
Because there is no modification to the nonlinear terms, there
remain exponentially growing solutions of the form
\begin{equation}
         F= ke^{\lambda t} - 1 -\frac{c}{\lambda}.
\label{eq:tf}
\end{equation}

Another example is the choice
\begin{equation}
  \hat \Gamma^\mu 
         = \frac {c}{\sqrt{-g}}(\gamma^{t\mu} -\gamma^{t\mu}_{[0]}) ,
\label{eq:ntf}
\end{equation}
where we use the notation $f_{[0]}=f(0,x^i)$.
In the spatially homogeneous case, this choice leads via the constraints
({\ref{eq:Gconstr}) to 
\begin{equation}
  \partial_t (\gamma^{t\mu}-\gamma_{[0]}^{t\mu})
      =- c( \gamma^{t\mu}-\gamma_{[0]}^{t\mu}),
\end{equation}
with solution $\gamma^{t\mu}=\gamma_{[0]}^{t\mu}$. Thus, if the constraints are
satisfied, $\gamma^{t\mu}$ is forced to retain to its initial value, which has
the potential advantage of warding off coordinate pathologies. If with the
gauge forcing term (\ref{eq:ntf}) we make the homogeneous Kerr-Schild ansatz
(\ref{eq:Fks}), then the reduced Einstein equations would require
\begin{equation}
  \hat E^{\mu\nu} 
       =\frac{1}{2}\bigg ( (1+F) F_{,tt} -F_{,t} F_{,t} +c F_{[0]}F_{,t} \bigg)
                       k^\mu k^\nu 
   + \frac{c}{2}F_{,t}\bigg(  \delta_y^\mu\delta_y^\nu
         + \delta_z^\mu\delta_z^\nu \bigg)   =0.
\end{equation}
Again the forcing term has only a linear effect but now it is inconsistent with
the Kerr-Schild ansatz (except in the trivial case $F_{,t}=0$), so that the
evolution would in general create other components in the metric and possibly
excite other instabilities.

Other gauge source functions can be chosen which introduce nonlinear terms  but
we have not found an example which preserves the Kerr-Schild form
(\ref{eq:Fks}). Consequently, an analytic analysis of their effect on the
shifted gauge wave is difficult to carry out. It is clear from black hole
simulations using harmonic coordinates that gauge forcing can play a helpful
role~\cite{pret1,pret2}, but there are few general guidelines to go by. In the
case of an analytic testbed, the addition of gauge forcing terms beyond those
specified in the analytic solution can complicate the test if the resulting
solution is not known. We have had no computational success in using gauge
forcing terms to control instabilities in the shifted gauge wave test.

\subsection{Constraint adjustment}
\label{sec:constadj}

There is no extensive knowledge regarding the stability of solutions
to the constraint systems (\ref{eq:constr}) or (\ref{eq:hatconstr}). 
Of special importance to numerical evolution is whether constraint violating
perturbations of the reduced Einstein equations can grow at a fast
rate. This question is complicated by the fact that the
reduced harmonic equations are not unique. They can be adjusted,
without affecting their hyperbolicity, according to
\begin{eqnarray}
   \tilde E^{\mu\nu} : &=&G^{\mu\nu} -\nabla^{(\mu} {\cal C}^{\nu)} 
                 +\frac{1}{2}g^{\mu\nu}\nabla_\alpha {\cal C}^\alpha
		 + A^{\mu\nu} \nonumber \\
                &=&E^{\mu\nu} +\nabla^{(\mu} \hat \Gamma^{\nu)} 
                 -\frac{1}{2}g^{\mu\nu}\nabla_\alpha \hat \Gamma^\alpha
		 + A^{\mu\nu}=0 
		 \label{eq:etilde}
\end{eqnarray}
provided the adjustment has the functional form 
\begin{equation}
         A^{\mu\nu}= A^{\mu\nu}
   (x^\alpha,g^{\alpha\beta},\partial_\gamma g^{\alpha\beta},{\cal C}^\alpha) 
\end{equation}
with $ A^{\mu\nu}=0$ when ${\cal C}^\alpha=0$.
The resulting constraint system (\ref{eq:hatconstr}) becomes
\begin{equation}
     \nabla^\alpha \nabla_\alpha {\cal C}^\mu +R^\mu_\nu {\cal C}^\nu 
                   -2\nabla_\nu A^{\mu\nu}=0,
     \label{eq:ahatconstr}
\end{equation}
with the reduced Ricci tensor given by
\begin{equation}
     R^{\mu\nu} = \nabla^{(\mu}{\cal C}^{\nu)}- A^{\mu\nu}
            +\frac{1}{2}g^{\mu\nu}A.
     \label{eq:aredricci}
\end{equation}
The standard form of the reduced harmonic equations (\ref{eq:e})
differ from Fock's~\cite{Fock} harmonic formulation, on which the original
version of the Abigel code was based~\cite{harl}, by the adjustment 
\begin{equation}
     A^{\mu\nu}=\frac{1}{2g}\Gamma^{(\mu}g^{\nu)\alpha}\partial_\alpha g
          -\frac{1}{4g}g^{\mu\nu}  \Gamma^\alpha \partial_\alpha g .
\label{eq:aold}
\end{equation}

In the absence of a general theory, computational experiments are necessary to
determine the effect of a given adjustment. However, some
clues can be provided by the following observations.

In the linear approximation, with unit lapse and zero shift,
the adjustment~\cite{constrdamp}
\begin{equation}
      A^{\mu\nu}= -\lambda {\cal C}^{(\mu}\nabla^{\nu)}t , \quad  \lambda>0  
\label{eq:constdamp}
\end{equation}
leads to the constraint system
\begin{equation}
        (-\partial_t^2 +\nabla^2){\cal C}^\mu =\lambda(\partial_t {\cal C}^\mu
	         +\delta^\mu_t \partial_\nu {\cal C}^\nu).
\end{equation}
The spatial components satisfy
\begin{equation}
        (-\partial_\tau^2 +\nabla^2)(e^{\lambda t} {\cal C}^i) =0,
	\label{eq:decay}
\end{equation}
where $\tau=(e^{\lambda t}-1)/\lambda$. For physically reasonable  boundary
conditions, e.g. periodic boundary conditions, a solution $F$ of the wave
equation (\ref{eq:decay}) must have finite energy, so that
\begin{equation}
       \int |{\bf \nabla} F|^2 dx dy dz <K_1
\end{equation}
for some constant $K_1$. Consequently, 
\begin{equation}
        \int |{\bf \nabla} C^i|^2 dx dy dz < K_1 e^{-2\lambda t}
\end{equation} which implies that any initial error in the constraint $C^i$
must decay to a spatially constant solution $C_\infty^i(t)$. Such homogeneous
solutions of (\ref{eq:decay}) have the form $C_\infty^i=K_2+K_3e^{-\lambda t}$.
By the analogous argument, any constraint violation in the time component
${\cal C}^t$ also decays to to a constant value. This result has previously
been established for the case $\partial_i {\cal C}^\mu\ne 0$ using mode
analysis~\cite{constrdamp}. 

Unfortunately, such constraint damping does not extend in a straightforward way
to the nonlinear case, where in particular it can lead to the excitation of
constraint preserving instabilities. For example, this adjustment does not
preserve the Kerr-Schild metric type (\ref{eq:Fks}) so that in damping the
instability (\ref{eq:instab}) it can excite an instability of the type
(\ref{eq:lgw}), which cannot be controlled by constraint adjustment. See
Fig.~\ref{fig:DDGaugeWave1Dsh} in Sec.~\ref{sec:tests} for evidence of this
behavior.

The exponential instability (\ref{eq:expt}) satisfies the reduced harmonic
equations but violates the harmonic constraints. Although we have found no way
to control this instability with harmonic gauge source functions $\hat
\Gamma^\mu$, it may be suppressed by the constraint adjustment
\begin{equation}
      A^{\mu\nu}= -\frac {c}{\sqrt{-g}} {\cal C}^\alpha 
              \partial_\alpha (\sqrt{-g}g^{\mu\nu}) , \quad  c>0 . 
\label{eq:beladj}
\end{equation}
The reduced equations (\ref{eq:etilde}) (with $\hat \Gamma^\mu =0$) now give
\begin{equation}
      2\sqrt{-g} \tilde E^{\mu\nu}= ( F_{,tt} -F_{,t} F_{,t} +FF_{,tt} 
             +2cF_t F_t)k^\mu k^\nu =0
\label{eq:cF}
\end{equation}
with solutions
\begin{equation}
       F=(a_1 t+ a_2)^{1/(2c)}-1.
\end{equation}
Thus the exponential growth has been removed.  For the  case $c=1/2$, in which
$F$ grows only linearly in time, the effect of this adjustment is to remove the
lowest differential order terms in ${\cal C}^\mu$ from (\ref{eq:etilde}), i.e.
\begin{equation}
       -\nabla^{(\mu} {\cal C}^{\nu)} 
               +\frac{1}{2}g^{\mu\nu}\nabla_\alpha {\cal C}^\alpha
	+ A^{\mu\nu}=\frac{1}{2}(-g^{\mu\alpha}\partial_\alpha {\cal C}^\nu
	     -g^{\nu\alpha}\partial_\alpha {\cal C}^\mu).
\end{equation}

The growth rate can be completely eliminated by the adjustment
\begin{equation}
      A^{\mu\nu}= -\frac{b}{2} g^{tt} k^{(\mu}{\cal C}^{\nu)}, \quad  b>0 .
\label{eq:mmbeladj}
\end{equation}
Then (\ref{eq:etilde}) gives
\begin{equation}
      2\sqrt{-g} \tilde E^{\mu\nu}= \bigg((1+F) (F_{,tt} +b F_{,t})
           -F_{,t} F_{,t} \bigg)k^\mu k^\nu =0
\label{eq:mmcF}
\end{equation}
with the {\em strongly} damped solution
\begin{equation}
      1+F=a_1 exp\bigg (a_2 e^{-bt}\bigg ) .
\label{eq:strdam}
\end{equation}
The same strongly damped behavior (\ref{eq:strdam}) follows from the
adjustment 
\begin{equation}
      A^{\mu\nu}= \frac{b {\cal C}^\alpha \nabla_\alpha t}
      {e_{\rho\sigma}{\cal C}^\rho {\cal C}^\sigma}
         {\cal C}^{\mu} {\cal C}^{\nu}, \quad  b>0 ,
\label{eq:mmmmbeladj}
\end{equation}
where 
\begin{equation}
        e_{\rho\sigma}=g_{\rho\sigma}- 
          \frac{2}{g^{tt}}(\nabla_\rho t)\nabla_\sigma t
\end{equation}
is the natural metric of signature $(++++)$ associated with the Cauchy
slicing. Here (\ref{eq:mmmmbeladj}) has the advantage over
(\ref{eq:mmbeladj}) that it has a geometric construction which is
independent of the Kerr-Schild form of the metric.

The adjustments (\ref{eq:beladj}) and especially (\ref{eq:mmmmbeladj}) lead to
improved performance in the shifted gauge wave tests described in
Sec~\ref{sec:tests}. Other adjustments can be designed to introduce lowest
differential order terms which act as a repulsive potential in the wave
equation (\ref{eq:ahatconstr}). For example, consider the adjustment
\begin{equation}
      A^{\mu\nu}= -k g^{\mu\nu} \Phi^2 , \, k>0 . 
\label{eq:phiadj}
\end{equation}
where $\Phi={\cal C}^\alpha {\cal C}_\alpha$
Then, after contraction with ${\cal C}_\mu$, (\ref{eq:ahatconstr}) gives
\begin{equation}
       \nabla_\alpha \nabla^\alpha \Phi -2k \Phi^3  ={\cal F},
\end{equation}
where ${\cal F}$ vanishes when $\nabla_\alpha {\cal C}^\beta =0$. By
choosing $k$ to be large, this repulsive
$\Phi^4$ potential might at the least constrain  ${\cal C}^\alpha$ to be
a null vector.

\section{Finite difference implementation}
\label{sec:imp}

A first differential order formalism is useful for applying the well
developed theory of symmetric hyperbolic systems but in a numerical code
it introduces extra variables and their associated nonphysical
constraints. For this reason, we base our code on the natural second
order form of the quasilinear wave equations which comprise the reduced
harmonic system (\ref{eq:etilde}). They are finite differenced in the
flux conservative form 
\begin{equation}
  2\sqrt{-g}\tilde E= \partial_\alpha(g^{\alpha\beta}\partial_\beta \gamma^{\mu\nu})-S^{\mu\nu} =0
    \label{eq:fc}
\end{equation}
where $S^{\mu\nu}$ is comprised of (nonlinear) first derivative
terms that do not enter the principle part.

Numerical evolution is implemented on a spatial
grid  $(x_{I}, y_{J}, z_{K})=(Ih, \, Jh, \, Kh)$, $0\le (I,J,K) \le N$,
with uniform spacing $h$, 
on which a field $f(t,x^i)$ is
represented by its grid values $f_{[I,J,K]}(t) = f(t, x_{I}, y_{J}, z_{K})$. 
The time integration is carried out by the method of lines using a 4th
order Runge-Kutta method. 
We introduce the standard finite difference operators $D_{0i}$
and $D_{\pm i}$ according to the examples 
\begin{eqnarray}
    D_{0x} f_{I,J}&=&\frac{1}{2h}(f_{I+1,J}-f_{I-1,J}) \nonumber \\
    D_{+x} f_{I,J}&=&\frac{1}{h}(f_{I+1,J}-f_{I,J}) \nonumber \\
   D_{-y} f_{I,J}&=&\frac{1}{h}(f_{I,J}-f_{I,J-1}) \nonumber \, ;
\end{eqnarray}
the translation operators $T_{\pm i}$ according to the
example 
\begin{equation}
        T_{\pm x}f_{I,J}=f_{I\pm 1,J}  
\end{equation}
and the averaging operators $A_{\pm i}$ and  $A_{0i}$ ,
according to the examples
\begin{eqnarray}
     A_{\pm x}f_{I,J} = \frac{1}{2}(T_{\pm x}+1)f_{I,J} \nonumber 
\end{eqnarray}
\begin{eqnarray}
     A_{0x}f_{I,J} = \frac{1}{2}(T_{+x}+T_{-x})f_{I,J}\nonumber .
\end{eqnarray}
Standard centered differences $D_{0i}$ are used to approximate the first
derivative terms in (\ref{eq:fc}) comprising $S^{\mu\nu}$.

We will describe the finite difference techniques for the principle part of
(\ref{eq:fc}) in terms of the scalar wave equation 
\begin{equation}
    \partial_\alpha(g^{\alpha\beta}\partial_\beta \Phi ) =0.
    \label{eq:scalarwave}
\end{equation}
Note that the principle part of the linearization of (\ref{eq:fc})  gives rise
to (\ref{eq:scalarwave}) for each component of $\gamma^{\mu\nu}$.  The
non-vanishing shift leads to mixed space-time derivatives
$\partial_t\partial_i$ which complicates the use of standard explicit
algorithms for the wave equation. This problem has been addressed
in~\cite{alcsch,calab,excis,calabgund}. Here we base our work on an evolution
algorithm shown to be stable for a model 1-dimensional wave equation with
shift~\cite{excis}. Provided $g^{ij}$ is positive definite, as is the case when
the shift is subluminal (i.e. when the evolution vector
$t^\alpha\partial_\alpha=\partial_t$ is timelike), this algorithm has the
summation by parts (SBP) property that gives rise to an energy estimate for
(\ref{eq:scalarwave}). SBP algorithms have proved effective in other numerical
relativity codes~\cite{live,discrenerg,sbp}.

The algorithm we use is designed to obey semi-discrete versions of two
conservation laws obeyed by (\ref{eq:scalarwave}).
These govern the monopole quantity
\begin{equation}
     Q=-\int_V g^{t\alpha}\Phi_{,\alpha} dV.
\label{eq:contQ}
\end{equation}
and the energy 
\begin{equation}
    E=\frac{1}{2}\int_V (-g^{tt}\Phi_{,t}^2 +g^{ij}\Phi_{,i} \Phi_{,j})dV,
\label{eq:contE}
\end{equation}
where $dV=dxdydz$. By assumption, the $t=const$ Cauchy hypersurfaces are
spacelike so that $-g^{tt}>0$. We also assume in the following that $g^{ij}$ is
positive definite (subluminal shift) so that $E$ provides a norm. Note that in
the gravitational case, where $\Phi$ represents $\gamma^{\mu\nu}$, there are 10
quantities $Q^{\alpha\beta}$ corresponding to (\ref{eq:contQ}), which have
monopole, dipole or quadrupole transformation properties depending on the
choice of indices.

For periodic boundary conditions (or in the absence of a boundary),
(\ref{eq:scalarwave}) implies strict monopole conservation $Q_{,t} =0$; and,
when the coefficients of the wave operator are frozen in time, i.e. when
$\partial_t g^{\alpha\beta}=0$, (\ref{eq:scalarwave}) implies strict
energy conservation $E_{,t}=0$. In the time-dependent, boundary-free case,
\begin{equation}
       E_{,t}=\frac{1}{2} \int_V ( g^{\alpha\beta}_{,t}\Phi_{,\alpha}
                          \Phi_{,\beta} ) dV.
\label{eq:et}
\end{equation}
which readily provides an estimate of the form
\begin{equation}
       E_{,t} < k E
\end{equation}
for some $k$ independent of the initial data for $\Phi$. Thus
the norm  is bounded relative to its initial value at $t=0$ by
\begin{equation}
       E < A E_0 e^{kt}.
\label{eq:est}
\end{equation}
The most restrictive value of $k$ depends upon the ratios
of the norms of the quadratic forms defined by the integrands
in (\ref{eq:contE}) and (\ref{eq:et}).

For the present purpose, it suffices to describe the finite difference
evolution algorithm in the 2-dimensional case with periodic boundary conditions
$f_{0,J}=f_{N,J}$, $f_{I,0}=f_{I,N}$. We define the semi-discrete versions of
(\ref{eq:contQ}) and (\ref{eq:contE})  as
\begin{equation}
   Q=h^2\sum_{(I,J)=1}^N (-g^{tt}\Phi_{,t} - g^{ti} D_{0i}\Phi).
\label{eq:perq}
\end{equation}
and
\begin{equation}
   E=h^2\sum_{(I,J)=1}^N{\cal E},
\label{eq:peren}
\end{equation}
where
\begin{eqnarray}
    {\cal E} = &-&\frac{1}{2}g^{tt}\Phi_{,t}^2
            +\frac{1}{4}(A_{+x}g^{xx})(D_{+x}\Phi)^2
             +\frac{1}{4}(A_{-x}g^{xx})(D_{-x}\Phi)^2 \nonumber \\
             &+&\frac{1}{4}(A_{+y}g^{yy})(D_{+y}\Phi)^2
             +\frac{1}{4}(A_{-y}g^{yy})(D_{-y}\Phi)^2+g^{xy}(D_{0x}\Phi)D_{0y}\Phi  .
\label{eq:caleB}    
\end{eqnarray}
The energy provides a norm on the discretized system, i.e. $E=0$
implies $\Phi_{,t}=D_{\pm i} \Phi=0$ (provided the grid spacing $h$ is
sufficiently small to justify regrouping of terms into perfect squares
as in the continuum case).

The simplest second order approximation to (\ref{eq:scalarwave}) which reduces
in the 1-dimensional case to the SBP algorithm presented in~\cite{excis} is 
\begin{equation}
   W:= - \partial_t(g^{tt} \partial_t\Phi) 
        -\partial_t(g^{ti}D_{0i}\Phi)
         - D_{0i}(g^{it} \partial_t\Phi)
                 -{\cal D}_g^2 \Phi =0
\label{eq:2wB}
\end{equation}
where
\begin{eqnarray}
      {\cal D}_g^2 \Phi  &=& 
            \frac{1}{2}D_{+x}\bigg((A_{-x}g^{xx}) D_{-x}\Phi \bigg)
          +\frac{1}{2}D_{-x}\bigg((A_{+x}g^{xx}) D_{+x}\Phi \bigg)
                                     \nonumber \\
         &+& \frac{1}{2}D_{+y}\bigg((A_{-y}g^{yy}) D_{-y}\Phi \bigg)
          +\frac{1}{2}D_{-y}\bigg((A_{+y}g^{yy}) D_{+y}\Phi \bigg)
                                     \nonumber \\
          &+& D_{0x}\bigg(g^{xy} D_{0y}\Phi \bigg)
           + D_{0y}\bigg(g^{xy} D_{0x}\Phi \bigg)  .   
\label{eq:cald}
\end{eqnarray}
It follows immediately from the flux conservative form of $W$ that $Q_{,t}=0$
for the case of periodic boundary conditions. In order to establish the SBP
property we consider the frozen coefficient case $\partial_t
g^{\alpha\beta}=0$. Then a straightforward calculation gives
\begin{eqnarray}
    {\cal E}_t-\Phi_t W &=& \frac{1}{2} D_{+i}\bigg( 
          g^{ti}\Phi_t T_{-i}\Phi_t+ \Phi_t T_{-i}(g^{ti}\Phi_t)
                                   \bigg)\nonumber \\
  &+&\frac{1}{2} D_{+x}\bigg( (A_{-x}g^{xx})(D_{-x}\Phi)T_{-x}\Phi_t \bigg)
    +\frac{1}{2} D_{-x}\bigg( (A_{+x}g^{xx})(D_{+x}\Phi)T_{+x}\Phi_t \bigg)
                   \nonumber \\
  &+&\frac{1}{2} D_{+y}\bigg( (A_{-y}g^{yy})(D_{-y}\Phi)T_{-y}\Phi_t \bigg)
   +\frac{1}{2} D_{-y}\bigg( (A_{+y}g^{yy})(D_{+y}\Phi)T_{+y}\Phi_t \bigg)
                   \nonumber \\   
   &+&\frac{1}{2} D_{+x}\bigg( \Phi_t T_{-x}(g^{xy}D_{0y}\Phi)
                     +g^{xy}(D_{0y}\Phi) T_{-x}\Phi_t \bigg)
                   \nonumber \\   
   &+&\frac{1}{2} D_{+y}\bigg( \Phi_tT_{-y}(g^{xy}D_{0x}\Phi)
                     +g^{xy}(D_{0x}\Phi)  T_{-y}\Phi_t \bigg) .
\label{eq:caledotB}    
\end{eqnarray}

Because each term in (\ref{eq:caledotB}) is a total $D_{\pm i}$, it follows
(for periodic boundary conditions) that $E_{,t}=0$ when $W=0$ is satisfied. When
the coefficients of the wave operator are time dependent, an energy estimate
can be established analogous to (\ref{eq:est}) for the continuum case.

We also consider a modification of the algorithm (\ref{eq:est}) by
introducing extra averaging operators according to
\begin{equation}
  \hat W := - \partial_t(g^{tt} \partial_t\Phi) 
        -\partial_t ( g^{ti}D_{0i}\Phi ) 
         -  D_{-i}\bigg( (A_{+i}g^{it})(A_{+i}\partial_t\Phi) \bigg)     
	    - \hat{D_g}^2 \Phi =0 ,
\label{eq:2wa}
\end{equation}
with
\begin{eqnarray}
      {\hat D}_g^2 \Phi  &=& 
            \frac{1}{2}D_{+x}\bigg((A_{-x}g^{xx}) D_{-x}\Phi \bigg)
          +\frac{1}{2}D_{-x}\bigg((A_{+x}g^{xx}) D_{+x}\Phi \bigg)
                                     \nonumber \\
         &+& \frac{1}{2}D_{+y}\bigg((A_{-y}g^{yy}) D_{-y}\Phi \bigg)
          +\frac{1}{2}D_{-y}\bigg((A_{+y}g^{yy}) D_{+y}\Phi \bigg)
                                     \nonumber \\
          &+& D_{-x}\bigg( (A_{+x}g^{xy}) (A_{+x}D_{0y}\Phi )\bigg)
           + D_{-y}\bigg((A_{+y}g^{xy})(A_{+y} D_{0x}\Phi) \bigg) .    
\label{eq:calda}
\end{eqnarray}
It is easy to verify that $\hat W=W+O(h^2)$ and both $W$ and $\hat W$ are
constructed from the same stencil of gridpoints. Although $\hat W$ does not
obey the exact SBP property with respect to the energy (\ref{eq:caleB}), the
experiments in  Sec.~\ref{sec:tests} show that it leads to significantly better
performance for the shifted gauge wave test.

For the time discretization, we apply the method of lines to the 
large system of ordinary differential equations
\begin{equation}
     {\bf \Phi}_{,tt}=\frac{1}{h}{\bf A \Phi}_{,t}+\frac{1}{h^2}{\bf B \Phi}.
\end{equation}
obtained from the spatial discretization.
Introducing 
\begin{equation}
   {\bf \Phi}_{,t}=\frac{1}{h}{\bf T}, 
\end{equation}
we obtain the first order system
\begin{equation}
\left(
\begin{array}{c}
{\bf T} \\
{\bf \Phi}    
\end{array}
\right)_t =  \frac{1}{h}
\left( 
\begin{array}{cc}
{\bf B} & {\bf A} \\
{\bf I} & {\bf 0}   
\end{array}
\right)  
\left( 
\begin{array}{c}
{\bf T} \\
{\bf V}    
\end{array}
\right) . 
\label{eq:fo}
\end{equation}
We solve the system numerically using a 4th order
Runge-Kutta time integrator. 

Dissipation can be added by modifying (\ref{eq:fo})
according to
\begin{eqnarray}
 {\bf T}_t\rightarrow  {\bf T}_t 
              + \epsilon_T h^3 {\cal D}_g^2 {\cal D}_g^2 {\bf T},  \nonumber \\
  {\bf \Phi}_t \rightarrow {\bf \Phi}_t
          +\epsilon_\Phi h^3 {\cal D}_g^2 {\cal D}_g^2 {\bf \Phi}.
\label{eq:diss}
\end{eqnarray}

The $W$ algorithm is unstable when $g^{ij}$ is not positive-definite or,
equivalently, when the evolution direction $t^\alpha$ is superluminal
(spacelike). There are alternative evolution algorithms for the second
differential order 1D wave equation which are stable in the superluminal
case~\cite{calab,excis}. These algorithms have important application to the
black hole excision problem in treating the region inside the event horizon.
However they have no advantage in the shifted gauge wave test because the
evolution is superluminal only for amplitudes $A>1$ for which the spacelike
nature of the Cauchy hypersurfaces breaks down. These alternative algorithms
remain stable in the subluminal case but they are not as accurate as the $W$
algorithm because they involve a wider stencil of grid points. Nevertheless, it
is useful to compare their accuracy with the $W$ algorithm. (See
Fig.~\ref{fig:WVGaugeWave1Dsh} in Sec.~\ref{sec:tests}.)

For that purpose we introduce the simplest generalization of the $V$ algorithm
considered in~\cite{excis} to the 3D case. It is related to the $W$
algorithm (\ref{eq:2wB}) by
\begin{equation}
   V=W + ({\cal D}_g^2 -{\cal D}_h^2))\Phi
   - {\cal D}_{\beta,2}^2 \Phi 
\label{eq:wv}
\end{equation}
where ${\cal D}_h^2$ is defined as in (\ref{eq:cald}) with $g^{ij}$
replaced by the spatial 3-metric
\begin{equation}
         h^{ij}=g^{ij}-\frac{ g^{ti}g^{tj} }{g^{tt}}.
\end{equation}
and the shift terms are finite differenced by
\begin{eqnarray}
   {\cal D}_{\beta,2}^2 \Phi=
	 D_{0x}\bigg ( \frac{ g^{tx}g^{tx} }{g^{tt}}D_{0x}\Phi \bigg )
	+ D_{0y}\bigg ( \frac{ g^{ty}g^{ty} }{g^{tt}}D_{0y}\Phi \bigg )
	+ D_{0x,2h}\bigg ( \frac{ g^{tx}g^{ty} }{g^{tt}}D_{0y,2h}\Phi \bigg )
        + D_{0y,2h}\bigg ( \frac{ g^{tx}g^{ty} }{g^{tt}}D_{0x,2h}\Phi \bigg ),
\label{eq:cdb}
\end{eqnarray}
with, for example, 
\begin{equation}
    D_{0x,2h} f(x,y) =\frac{ f(x+2h,y)-f(x-2h,y)}{4h}
        =(1+\frac{h^2}{2} D_{+x}D_{-x})D_{0x} f.
\end{equation}
Equivalently, $V$ and $W$ are related by the $O(h^2)$ terms
\begin{eqnarray}
  V &=& W-\frac {h^2}{4} \bigg (
         D_{+x} D_{-x} \frac{ g^{tx}g^{tx} }{g^{tt}}D_{+x} D_{-x}
	+ D_{+y} D_{-y} \frac{ g^{ty}g^{ty} }{g^{tt}}D_{+y} D_{-y} 
	 \bigg ) \Phi \nonumber \\
     &-&\frac {h^2}{2} \bigg (
          D_{0x,2h}\frac{ g^{tx}g^{ty}}{g^{tt} } D_{+y} D_{-y}D_{0y}
	 +D_{0y,2h}\frac{ g^{tx}g^{ty}} {g^{tt}}D_{+x} D_{-x}D_{0x} \nonumber \\
       &+&D_{+x} D_{-x}D_{0x}\frac{ g^{tx}g^{ty}}{g^{tt} }D_{0y}  
	 +D_{+y} D_{-y}D_{0y}\frac{ g^{tx}g^{ty}}{g^{tt} }D_{0x}
       \bigg ) \Phi .
\label{eq:awv}
\end{eqnarray}
We also consider the modification
\begin{equation}
   \hat V=\hat W + ({\cal D}_g^2 -{\cal D}_h^2)\Phi
   - {\cal D}_{\beta,2}^2 \Phi .
\label{eq:hatwv}
\end{equation}

\section{Tests of the evolution algorithm}
\label{sec:tests}

We test the evolution code using the gauge wave and shifted gauge wave testbeds
with periodic boundary conditions and amplitude $A=.5$.  Test results for the
standard AwA gauge wave with amplitudes $A=.01$  and $A=.1$ using an earlier
version of the Abigel code can be found at the Pitt numerical relativity web
site~\cite{pitt}. (Those tests were performed using an  iterated
Crank-Nicholson time integrator on a code similar to the ${\hat W}$ algorithm
(\ref{eq:2wa}) with the constraint adjustments (\ref{eq:aold}) and
(\ref{eq:beladj}) with $c=.5$). The present code shows similar performance for
the standard AwA gauge wave test.

The tests are run on grids with $N=50 \rho$ zones, where $\rho= (1,2,4)$.
We use the $\ell_\infty$ norm to measure the error
\begin{equation}
       {\cal E}=||\Phi_\rho-\Phi_{ana}||_\infty
\end{equation}
in a grid function $\Phi_\rho$ with known analytic value $\Phi_{ana}$.
We measure the convergence rate at time $t$
\begin{equation}
    r(t) = \log_2 \big (
   \frac{||\Phi_2-\Phi_{ana}||}{||\Phi_4-\Phi_{ana}||} \big ),
\end{equation}
using the $\rho=2$ and $\rho=4$ grids ($N=100$ and $N=200)$.
It is also convenient to graph the rescaled error
\begin{equation}
      {\cal E_\rho} =\frac{\rho^2}{16}||\Phi_\rho-\Phi_{ana}||_\infty,
\end{equation}
which is normalized to the $\rho=4$ grid.

\subsection {Tests for the gauge wave without shift}

Figures~\ref{fig:GaugeWave1D} and~\ref{fig:GaugeWave2D} plot our results for
the  rescaled error ${\cal E_\rho}$ in $g_{xx}$, for $\rho =1,2,4$
($N=50,100,200$), for the 1D and 2D gauge wave tests with the gauge wave with
amplitude $A=.5$ using the bare $\hat W$ algorithm (\ref{eq:2wa}) (no
additional gauge forcing, constraint adjustment or dissipation). The 1D runs
were stopped at $t=1000$. At that time the absolute error for the coarsest grid
was unacceptably large  (over 100\%) but the error for the $\rho=4$ grid was
only $\sim 20$\%.  We found the convergence rates
\begin{equation}
       {r(50)} = 2.019, \quad {r(500)} = 1.677
\end{equation}
for the error in $g_{xx}$, measured at t=50 and t=500 (corresponding to 50 and
500 grid crossing times). For computational economy, the 2D runs, where the
wave propagates along a diagonal in the $(x,y)$ plane, were stopped at $t=100$.
We measured the convergence rates
\begin{equation}
       {r(10)} = 2.084, \quad  {r(100)} = 1.568
\end{equation}
at $t=10$  and $t=100$. Tests runs with the $W$ algorithm (\ref{eq:2wB}) showed
similar performance  although for the  coarsest grid the ${\hat W}$ algorithm
had noticeably smaller error, presumably as a result of the extra averaging
operators.

\begin{figure}[hbtp]
  \centering
  \includegraphics*[width=8cm]{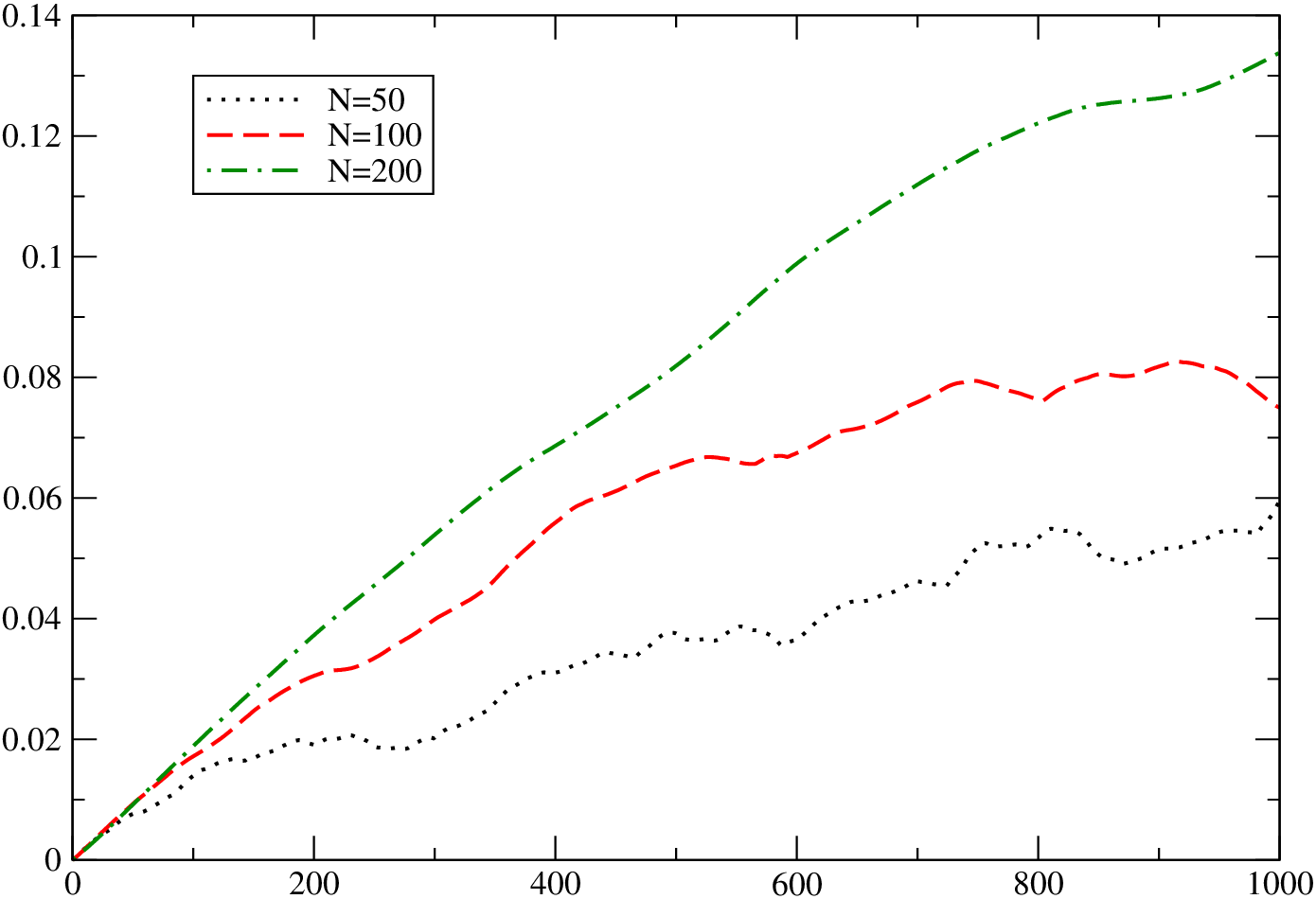}
  \caption{Plot of the rescaled error ${\cal E}_\rho(t)$ for the 1D gauge wave.}
  \label{fig:GaugeWave1D}
\end{figure}
 
\begin{figure}[hbtp]
  \centering
  \includegraphics*[width=8cm]{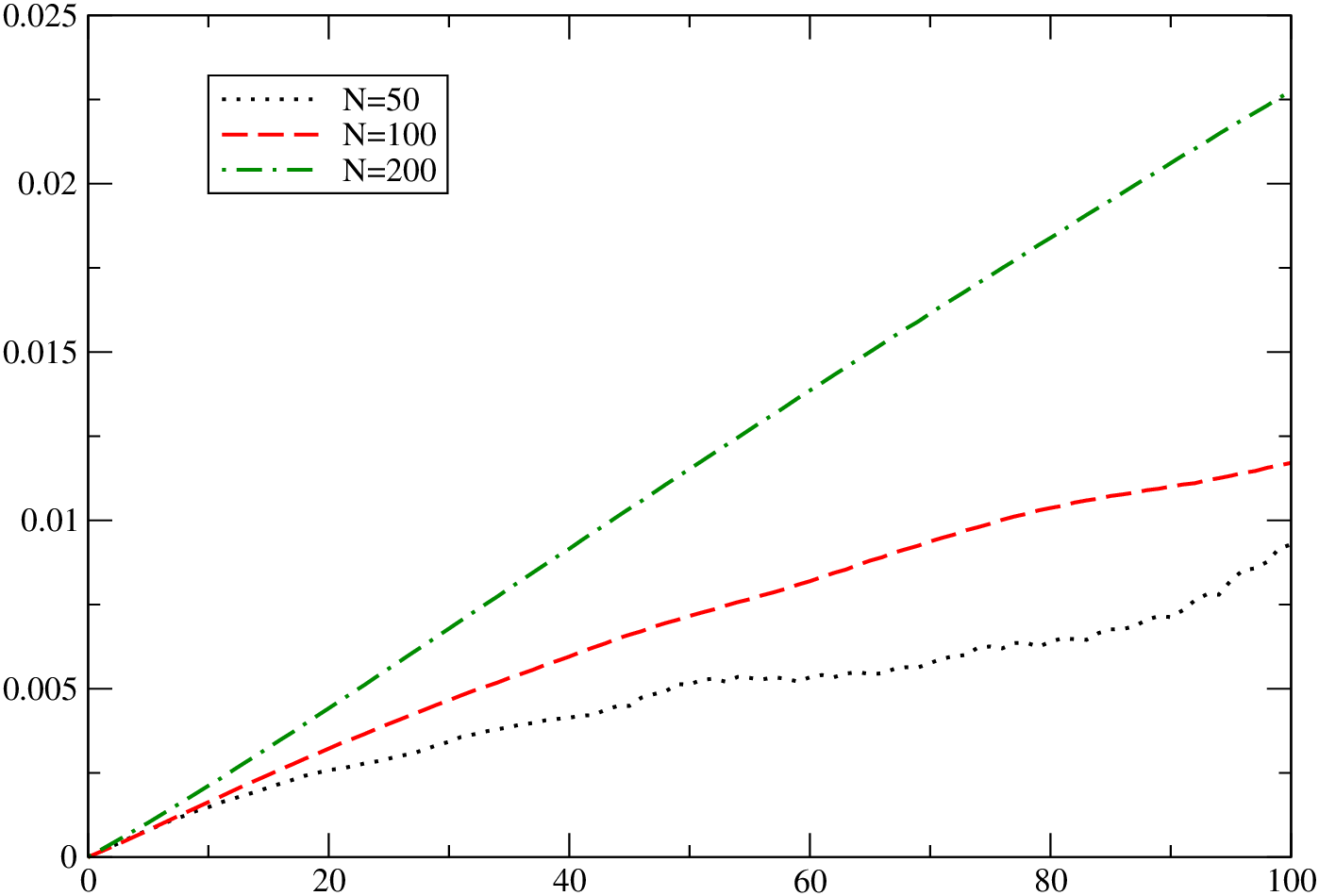}
  \caption{Plot of the rescaled error ${\cal E}_\rho(t)$ for the 2D gauge wave.}
  \label{fig:GaugeWave2D}
\end{figure}

The bare $W$ or $\hat W$ algorithms give excellent results for these gauge wave
tests. No appreciable improvement is attained by using the gauge source
functions or constraint adjustments discussed in Sec.~\ref{sec:cauchy} or by
using dissipation. It should be emphasized that this success is due to the
discrete conservation laws built into the algorithm. In comparison,
Fig.~\ref{fig:GaugeWave1D_gtt} shows snapshots of $g_{tt}(x)$ for a simulation 
of the gauge wave with $A=.5$ using an earlier version of the code without
these conservative properties. The snapshots are almost perfect matches to the
exponentially unstable mode (\ref{eq:lgw}).
Table~\ref{tab:GaugeWave} summarizes our results.

\clearpage
\begin{figure}[hbtp]
  \centering
  \psfrag{xlabel}{x}
  \psfrag{ylabel}{$g_{tt}$}
  \includegraphics*[width=8cm]{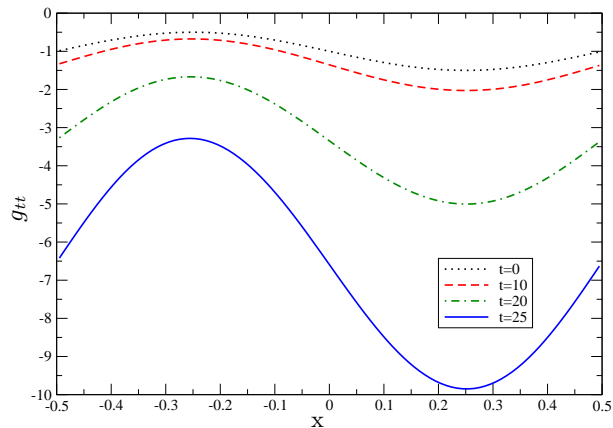}
  \caption{Snapshots of $g_{tt}(x)$ for the 1D gauge wave without shift
    obtained with an earlier version of the code without conservative
    properties, on an $N=100$ grid. The error almost exactly matches the
    unstable mode (\ref{eq:lgw}).
    }
  \label{fig:GaugeWave1D_gtt}
\end{figure}

\begin{table}[htbp]
  \begin{center}
    \begin{tabular}[c]{|c|c|}
\hline
   Test &
   Performance 
    \\ \hline \hline
 Non-conservative algorithm, 1D & Exponentially unstable mode (Fig.~\ref{fig:GaugeWave1D_gtt})
\\ \hline
 $W$ versus $\hat W$ algorithm, 1D & $\hat W$ has smaller error, due to extra averaging operators
\\ \hline
 $W$ versus $\hat W$ algorithm, 2D & $\hat W$ has smaller error
\\ \hline
$\hat W$ algorithm, convergence, 1D & Long term second order convergence (Fig.~\ref{fig:GaugeWave1D})
\\ \hline
$\hat W$ algorithm, convergence, 2D & Long term second order convergence (Fig.~\ref{fig:GaugeWave2D})
\\ \hline
\end{tabular}
    \caption{ Summary of tests performed 
              for the gauge wave without shift, with amplitude $A=0.5$. 
              Gauge forcing terms, constraint adjustments and dissipation 
              all gave no appreciable improvement.
            }
    \label{tab:GaugeWave}
  \end{center}
\end{table}


\subsection{Tests for the Shifted Gauge Wave}

For the shifted gauge wave tests we first consider the bare algorithms
(no additional constraint gauge forcing, adjustment or dissipation). In this
case, the $\hat W$ algorithm shows marked improvement over the $W$ algorithm. 
This is illustrated in Fig.~\ref{fig:WhWGaugeWave1Dsh} for the 1D wave with
A=.5, where the ${\hat W}$ algorithm runs more than twice as long as the $W$
algorithm with equal error.  For that reason we confine our attention to the
$\hat W$ algorithm in the remaining tests.
With the ${\hat W}$ algorithm, at $t=50$ we measured a convergence rate of
\begin{equation}
       r(50) = 2.135
\end{equation}
in the error of $g_{xx}$ for the shifted gauge  wave. The error is plotted in
Fig.~\ref{fig:hWGaugeWave1Dsh} for grids with $N=$ $50$, $100$ and $200$.

\begin{figure}[hbtp]
  \centering
   \includegraphics*[width=8cm]{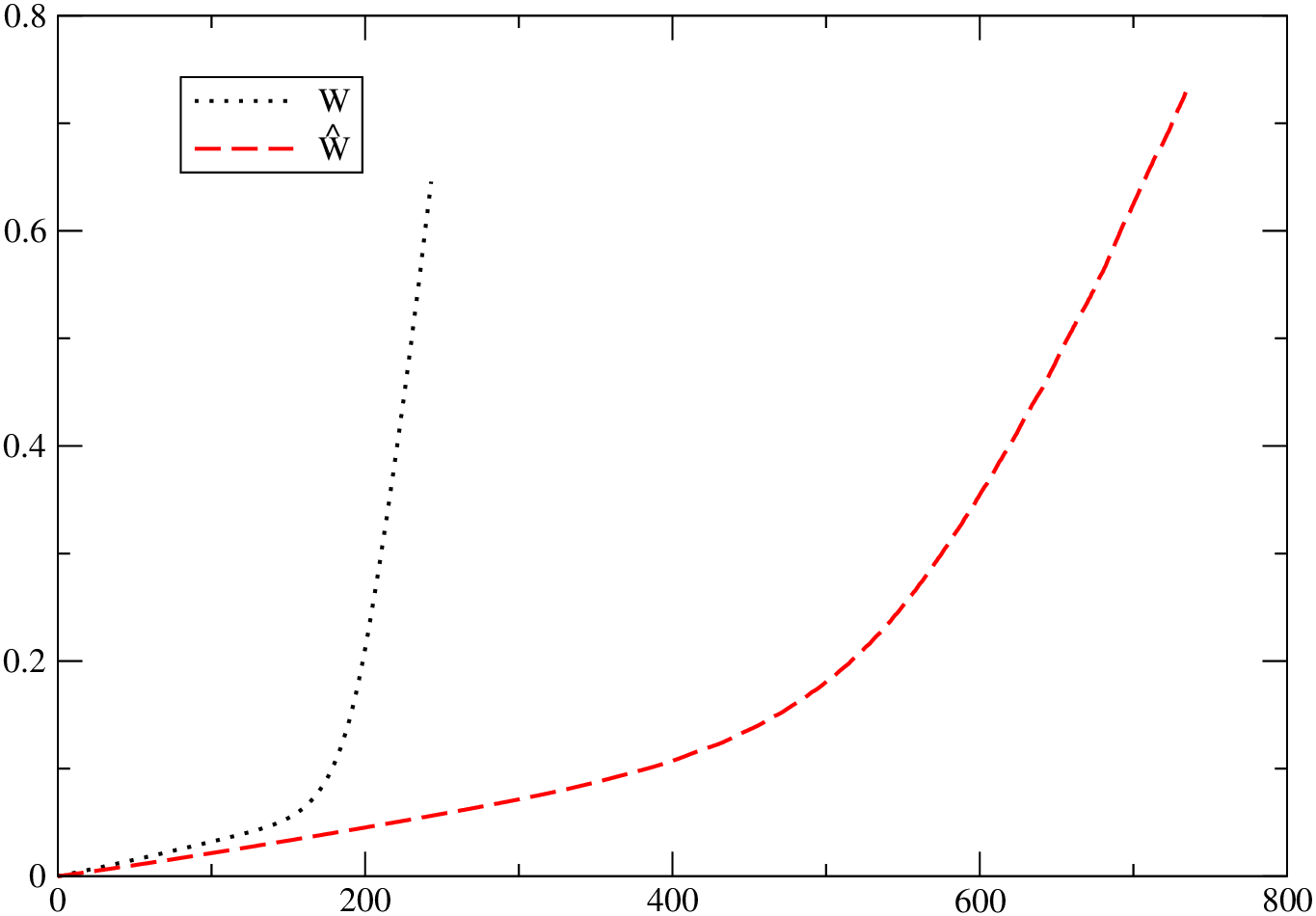}
  \caption{Comparison plots of the unscaled error ${\cal E}(t)$ in $g_{xx}$ 
 obtained with the bare $W$ and $\hat W$ algorithms
 for the 1D shifted gauge wave test, on an N=200 grid. Although the $W$
 algorithm has the SBP property, the additional averaging operators in the
 $\hat W$ algorithm lead to increased accuracy in this nonlinear test.}
  \label{fig:WhWGaugeWave1Dsh}
\end{figure}

\begin{figure}[hbtp]
  \centering
  \includegraphics*[width=8cm]{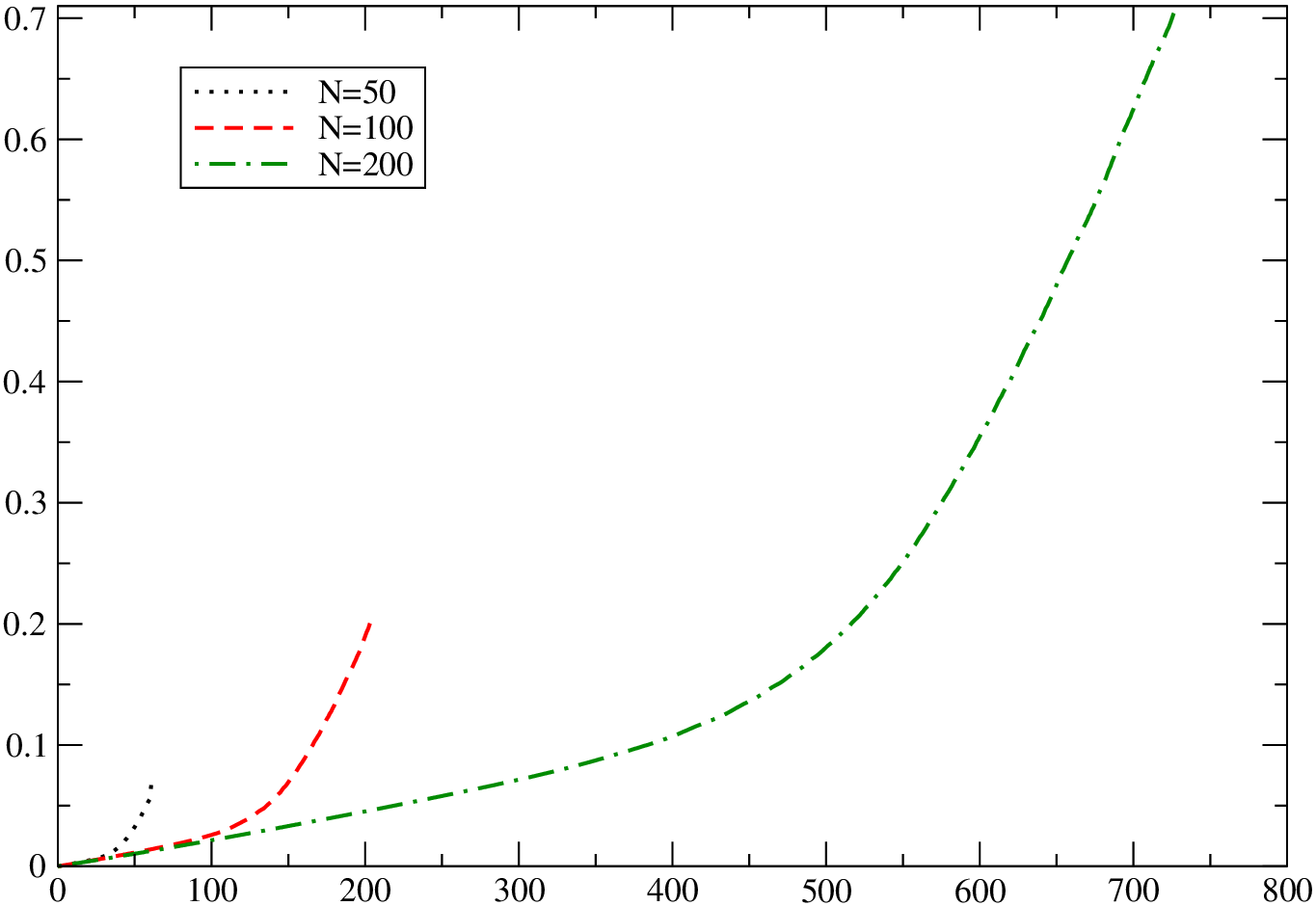}
  \caption{Plot of the rescaled error ${\cal E}_\rho(t)$ in $g_{xx}$
   obtained with the
   bare $\hat W$ algorithm for the 1D shifted gauge wave test.}
  \label{fig:hWGaugeWave1Dsh}
\end{figure}

The major error in the simulations with the bare algorithms is in the long
wavelength exponential mode (\ref{eq:expt}), which cause the runs with the
coarser grids to crash at an early time. This  is evident from the series of
snapshots of $g_{tt}(x)$ shown in Fig.~\ref{fig:GaugeWave1Dsh_gtt} for the
$N=100$ grid. At $t\approx 200$, the long wavelength error triggers the
superluminal instability of the $W$ algorithm as $g_{tt}\rightarrow 0$ at its
peak value. The snapshots of $g_{tt}(x)$ closely match the unstable constraint
violating mode (\ref{eq:expt}). The exponential term in (\ref{eq:expt})
introduces a positive error in $g_{tt}$ but a negative error in $g^{tt}$ so
that the Cauchy hypersurfaces remain spacelike as $g_{tt}\rightarrow 0$. 
As a result, this error does not trigger an instability in the $V$ algorithm
which is stable in the superluminal case, as discussed in Sec.~\ref{sec:imp}.

\begin{figure}[hbtp]
  \centering
  \includegraphics*[width=8cm]{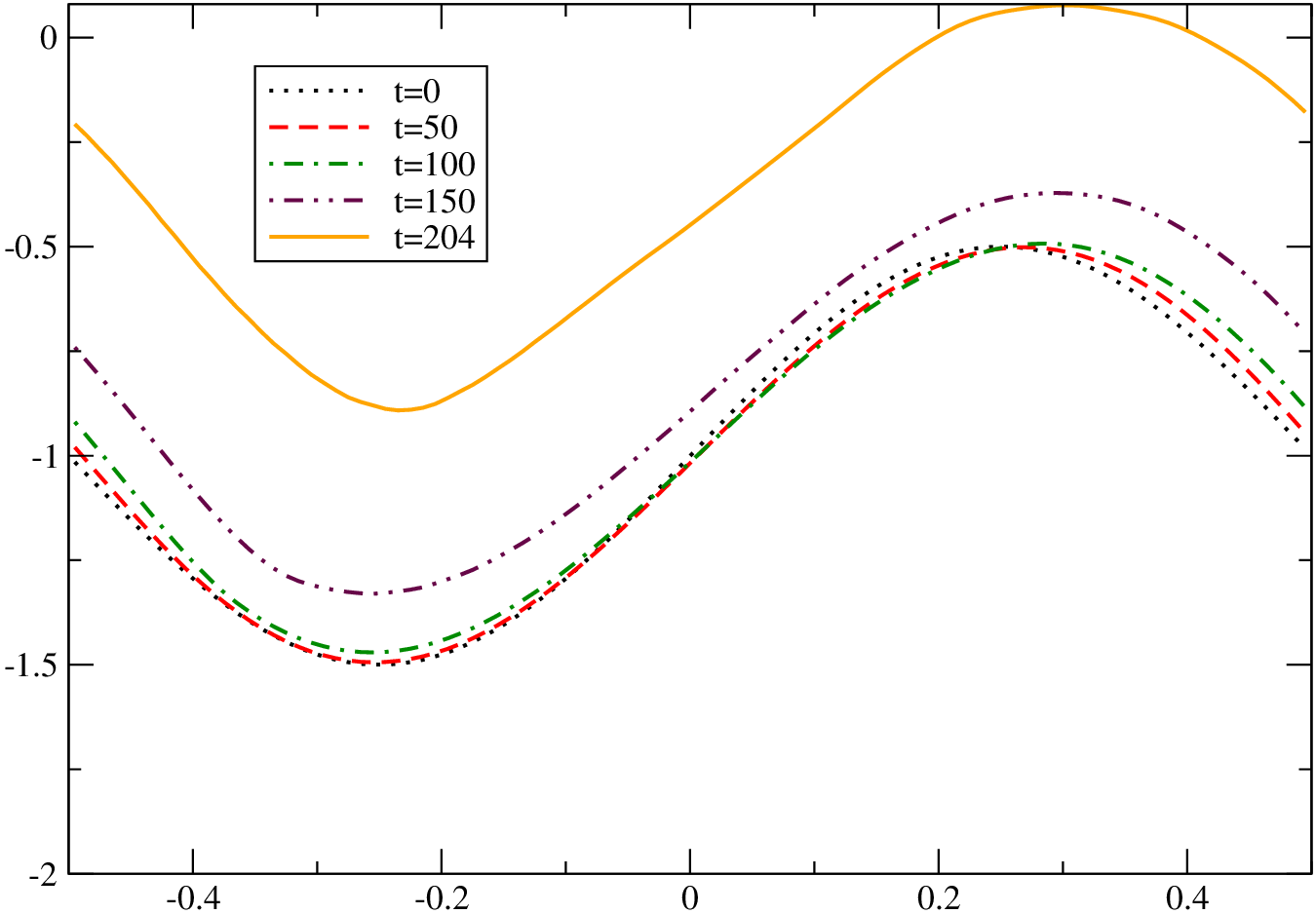}
  \caption{Snapshots of $g_{tt}(x)$ of the shifted gauge wave
        obtained with the bare ${\hat W}$ algorithm on an $N=100$ grid.
        The behavior closely matches the unstable mode
        (\ref{eq:expt}). At $t\approx 200$, the superluminal instability of
        the $W$ algorithm is triggered.
    }
  \label{fig:GaugeWave1Dsh_gtt}
\end{figure}

Tests of the bare $V$ and $\hat V$ algorithms (\ref{eq:wv}) and
(\ref{eq:hatwv}) with the 1D shifted gauge wave gave convergent results but the
error was considerably larger than with the bare $W$ and $\hat W$ algorithms.
Fig.~\ref{fig:WVGaugeWave1Dsh} compares the $\ell_\infty$ error in $g_{xx}$
between the $\hat V$ and the $\hat W$ algorithms. The $\hat V$ algorithm
remains stable well beyond the time $t\approx 200$ when the error brings it
into the superluminal regime. However, at $t\approx 200$ the error in the
simulation is $\approx 100 \%$  and has begun to grow rapidly, while the error
in the $\hat W$ algorithm is still small. Thus the trade-off for the stability
of the $V$ algorithm in the superluminal regime is its relative inaccuracy in
the subluminal regime compared to the $W$ algorithm. These results support the
strategy adopted in~\cite{excis} for a model black hole excision problem, in
which the $V$-algorithm is used in the inner region containing the horizon and
the $W$-algorithm is used in an outer region extending to the outer
boundary, with the two algorithms blended in a transition region.

\begin{figure}[hbtp]
  \centering
  \includegraphics*[width=8cm]{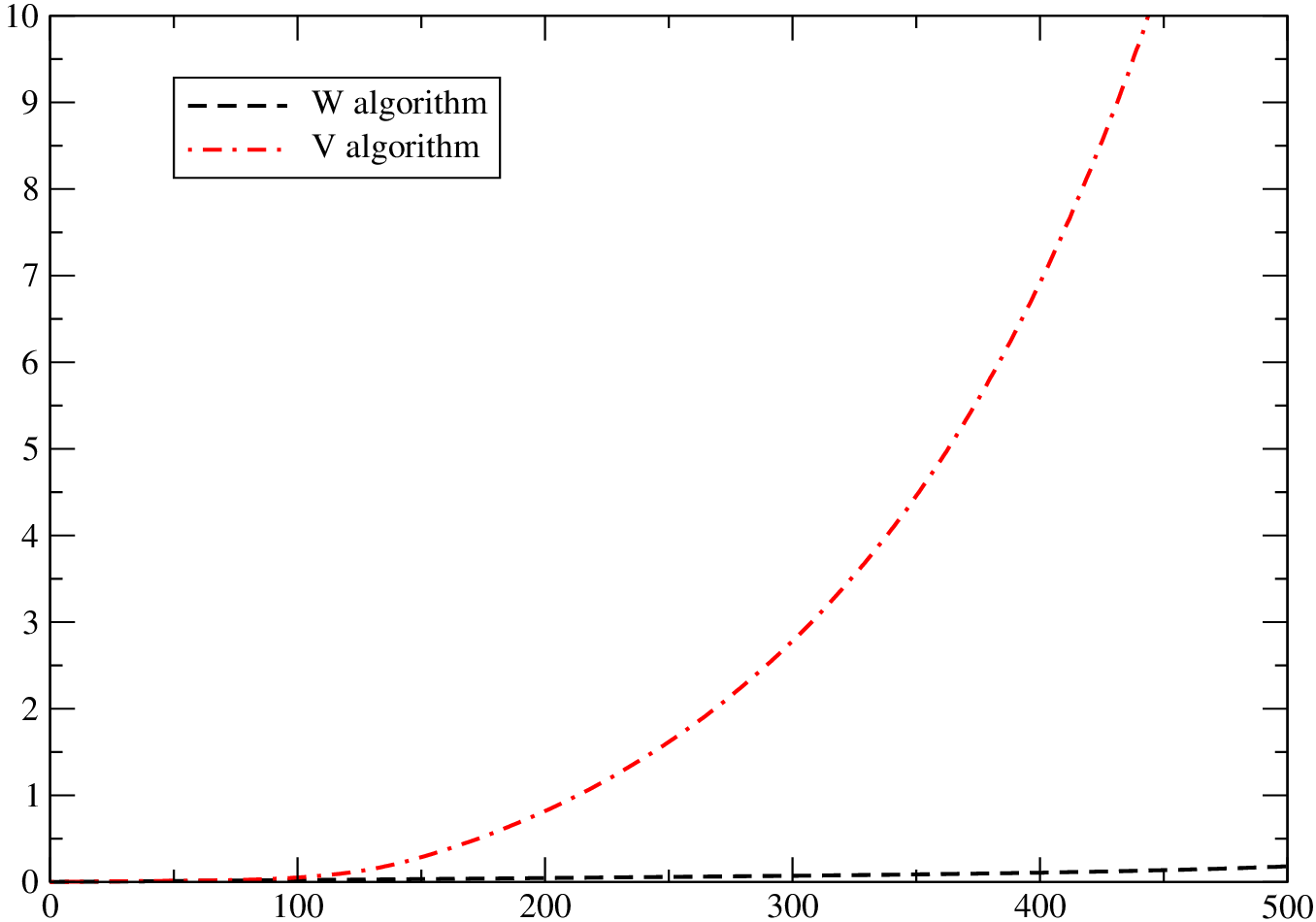}
  \caption{ Plots of the unscaled error ${\cal E}(t)$ in $g_{xx}$
   on an $N=200$ grid comparing the bare  $\hat W$ and $\hat V$ algorithms
   for the 1D shifted gauge wave.
	}
  \label{fig:WVGaugeWave1Dsh}
\end{figure}

We next consider tests with constraint adjustment. Excellent
performance of the $\hat W$ algorithm results from the constraint adjustment
(\ref{eq:beladj}). The error in $g_{xx}$ for the $N=100$ grid is plotted in
Fig.~\ref{fig:CGaugeWave1Dsh} for this adjustment with  $c=0$, $c=.5$ and
$c=1$. As evident from the plots, the $c=1$ run remains under control at
$t=1000$ whereas the unadjusted run crashes at $t \approx 200$. This
adjustment with $c=0.5$ also leads to improvement over the unadjusted case,
but the results are much more modest. 

\begin{figure}[hbtp]
  \centering
  \includegraphics*[width=8cm]{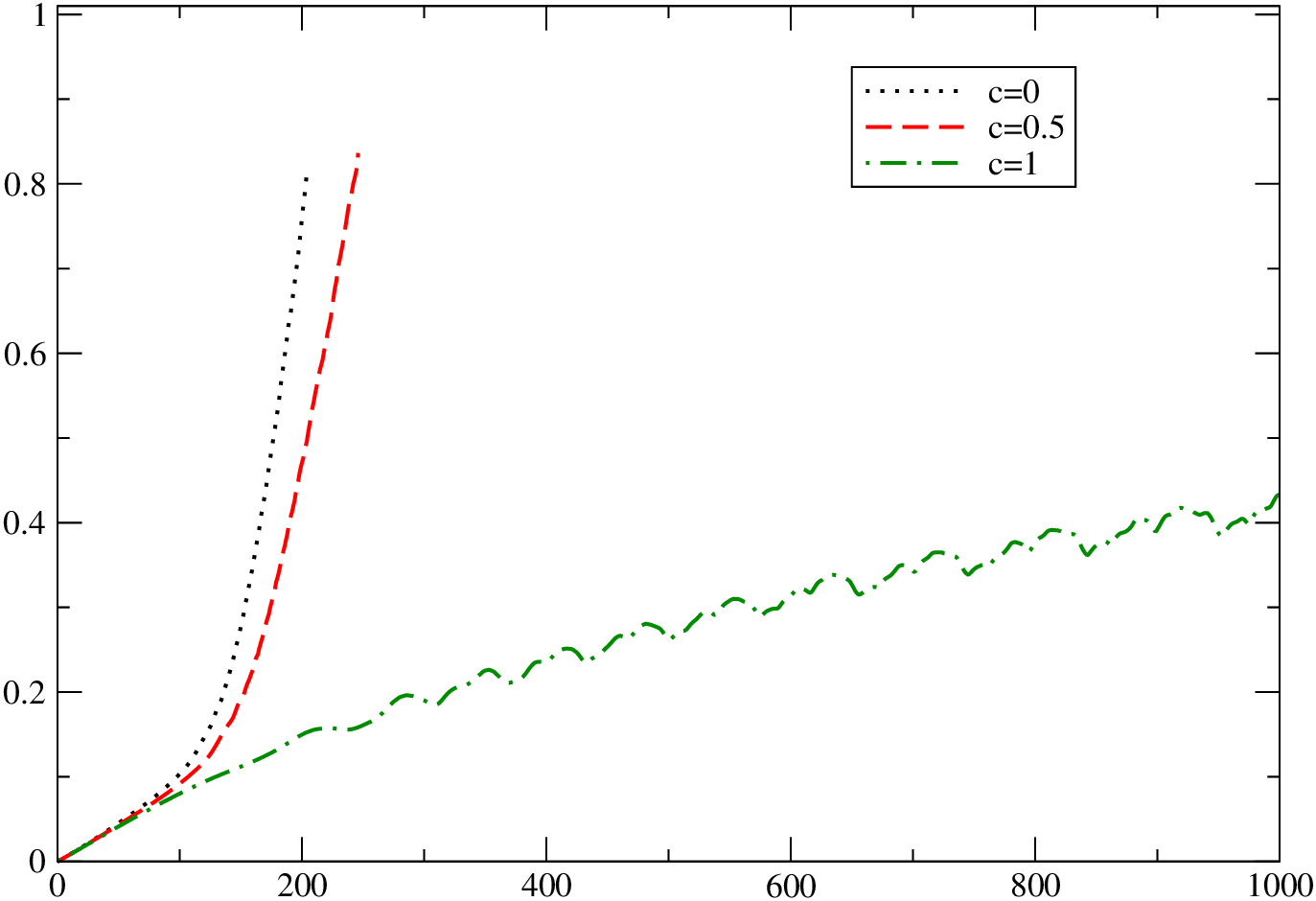}
  \caption{The unscaled error ${\cal E}(t)$ in the component $g_{xx}$,
   on an $N=100$ grid, using the
   constraint adjustment (\ref{eq:beladj}) in the 1D shifted gauge wave test.
   The adjustment with $c=1$ very effectively suppresses the instability. }
  \label{fig:CGaugeWave1Dsh}
\end{figure}

The constraint damping adjustment (\ref{eq:constdamp}) leads to some
improvement over the unadjusted case but the results are not nearly as
effective as the adjustment (\ref{eq:beladj}). The analysis leading to the
decaying behavior (\ref{eq:decay}) in the linear regime does not lead to
substantially improved performance in this nonlinear test. As shown in
Fig.~\ref{fig:DDGaugeWave1Dsh}, the error obtained with the constraint damped
$\hat W$ algorithm, with $\lambda=1$, is slightly smaller at early times than
for the undamped case. However, the error for the damped case then goes through
large oscillations and is roughly the same as for the undamped case at $t=200$.
Although the constraint damped case runs longer, the highly oscillatory
behavior produces unacceptably large error. The constraints ${\cal C}^\mu$
exhibit similar oscillation, indicating a coupling to a constraint preserving
mode. Our experiments indicate significant improvement cannot be obtained by
choosing other values of the damping coefficient $\lambda$ or by replacing
$t^\alpha$ in (\ref{eq:constdamp}) by another timelike vector, such as the
vector $\nabla^\alpha t$ normal to the Cauchy hypersurfaces. (We obtained
slightly better results for $t^\alpha$.) The addition of numerical dissipation
(\ref{eq:diss}) prolongs the run but does not control the large oscillations
produced by the constraint damping and does not lead to any substantial
increase in accuracy. Fig.~\ref{fig:DDGaugeWave1Dsh} shows the effect of
dissipation with $\epsilon_\Phi =.1$.

\begin{figure}[hbtp]
  \centering
  \includegraphics*[width=8cm]{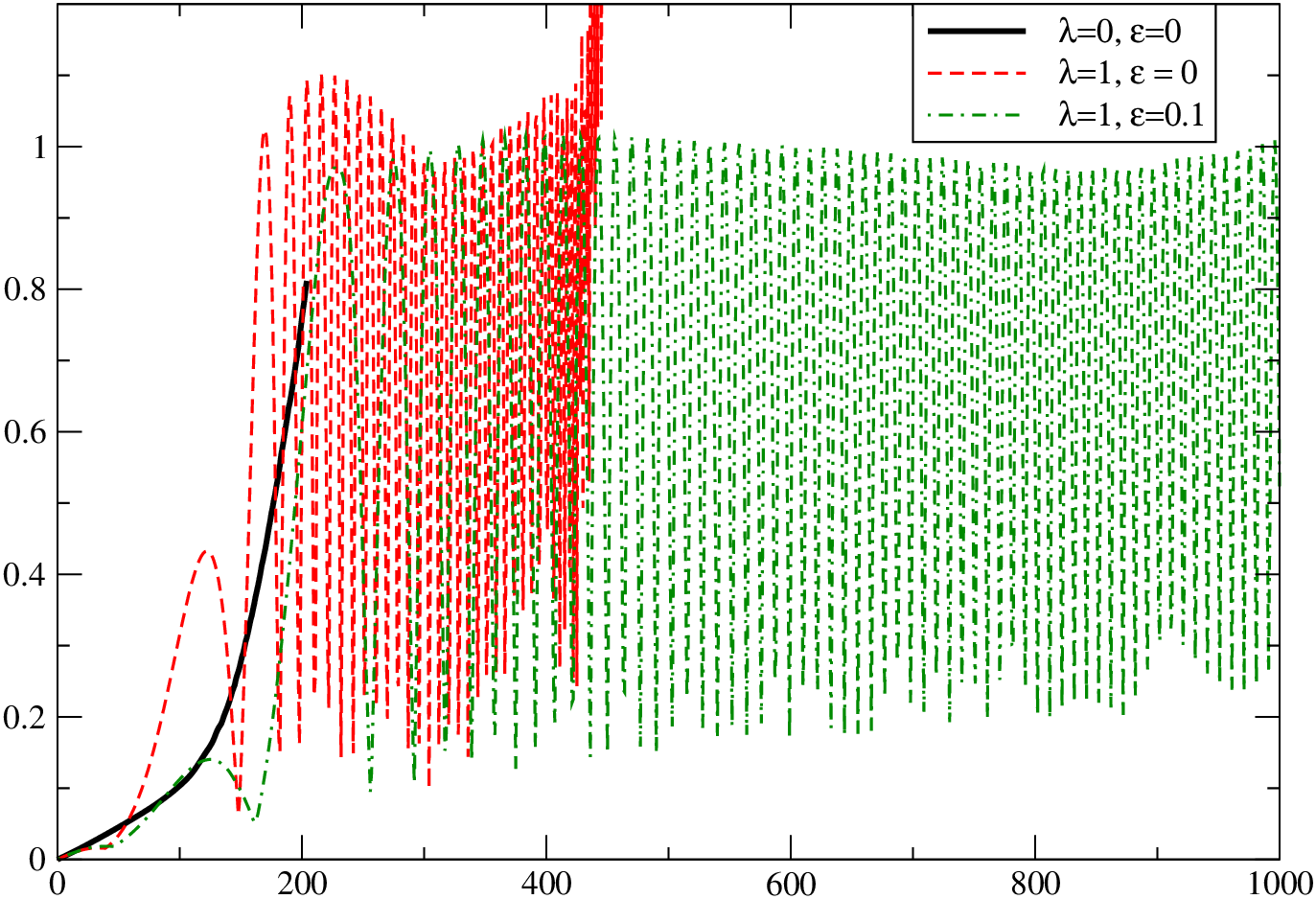}
  \caption{The effect of constraint damping ($\lambda=1$) and
	dissipation ($\epsilon=.1$) for the 1D shifted gauge wave test, on an
        $N=100$ grid. The graph of the unscaled error ${\cal E}(t)$ in $g_{xx}$
        shows a strong oscillation introduced by constraint damping. Although
        the simulation is extended by constraint damping and
        extended even more by dissipation,
	the duration of good accuracy is the same as with the bare
	algorithm. The constraints ${\cal C}^\mu$ exhibit similar oscillation,
        indicating coupling of the error to a constraint preserving mode.
   }
  \label{fig:DDGaugeWave1Dsh}
\end{figure}

In addition to the $c$-adjustment (\ref{eq:beladj}), we found that the
$b$-adjustment (\ref{eq:mmmmbeladj}) also gives excellent performance in the
shifted gauge wave test. Fig.~\ref{fig:GaugeWave1Dsh_new} compares the error,
on an $N=200$ grid, obtained using the ${\hat W}$ algorithm when adjusted by
(\ref{eq:mmmmbeladj}) with $b=1$ and when adjusted by (\ref{eq:beladj}) with
$c=1$.  Both adjustments are effective in controlling the unstable mode, which
is evident in the graph for $b=c=0$. The error for the two constraint adjusted
cases looks very similar, although a closer inspection shows small oscillations
in the $c=1$ graph, of the sort evident in Fig.~\ref{fig:CGaugeWave1Dsh} on an
$N=100$ grid, but which do not appear for $b=1$.

\begin{figure}[hbtp]
  \centering
  \includegraphics*[width=8cm]{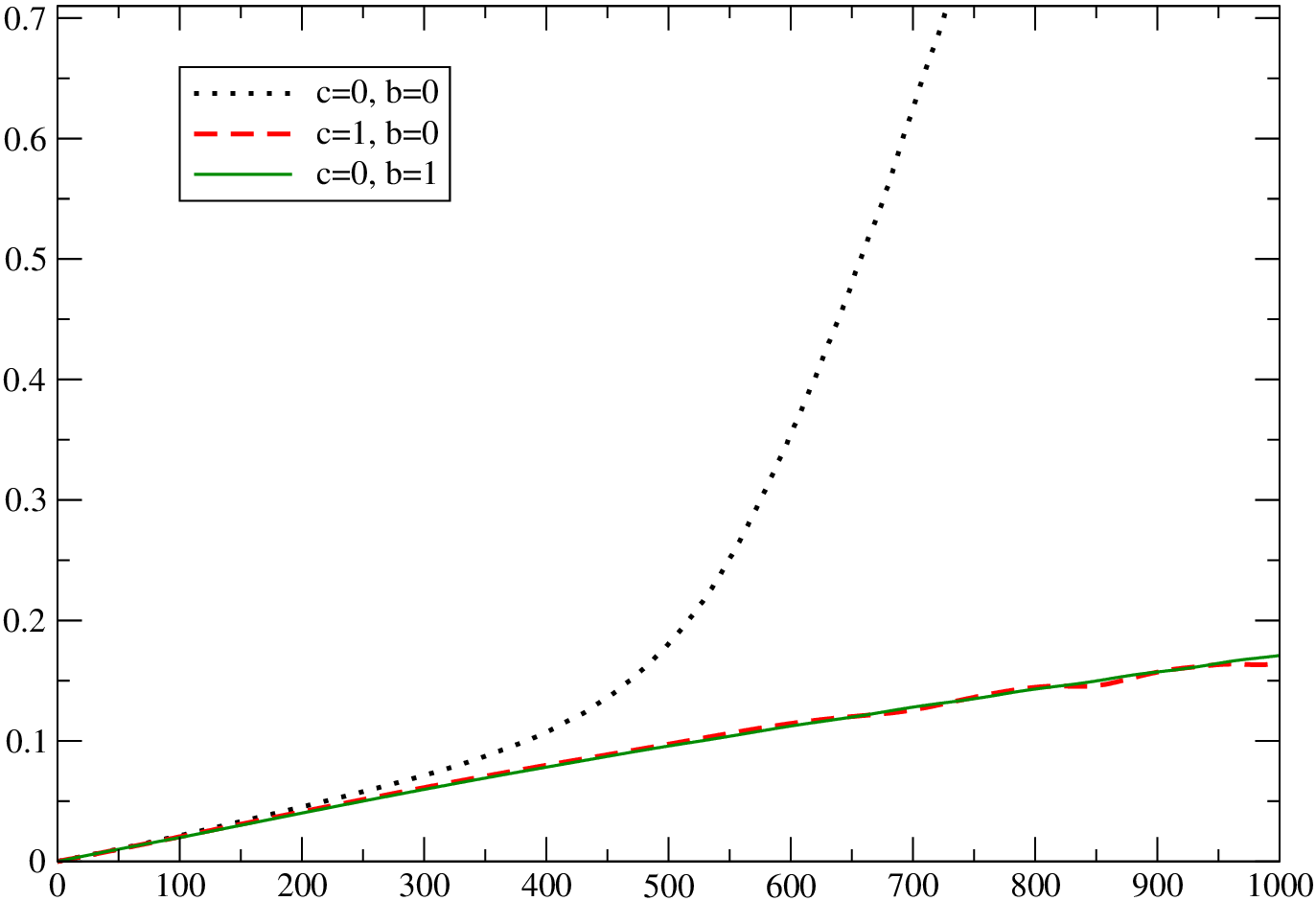}
  \caption{Comparison plots of the unscaled error ${\cal E}(t)$ in $g_{xx}$
       on an $N=200$ grid for the 1D shifted gauge wave test using the
       $\hat W$ algorithm with the constraint adjustment (\ref{eq:mmmmbeladj})
       with $b=1$, the constraint adjustment (\ref{eq:beladj}) with $c=1$ and
       the bare algorithm. The two adjustments show very similar error
       and both give excellent suppression
       of the unstable mode excited by the bare algorithm.
     } 
  \label{fig:GaugeWave1Dsh_new}
\end{figure}

Our experiments with gauge forcing terms, such as (\ref{eq:ntf}), gave no
significant improvement in performance. However, this is likely a feature of the
shifted gauge wave test for which harmonic coordinates do not encounter any
pathologies. As reported in~\cite{pret1,pret2}, we expect gauge forcing terms
to be essential for the long term simulation of black holes.  

We found the 1D shifted gauge wave tests to be very effective in determining
those algorithms which would give good performance in the 2D test. For that
reason, we limit our presentation of 2D test results for the $\hat W$ algorithm
with the $b=1$ constraint adjustment (\ref{eq:mmmmbeladj}).
Fig.~\ref{fig:GaugeWave2Dsh_new} shows the rescaled error obtained using
$N=50$, $N=100$ and $N=200$ grids. The convergence rate $r(100)=2.050$ was
measured at $t=100$. Table~\ref{tab:GaugeWavesh} summarizes our results.

\begin{figure}[hbtp]
  \centering
  \includegraphics*[width=8cm]{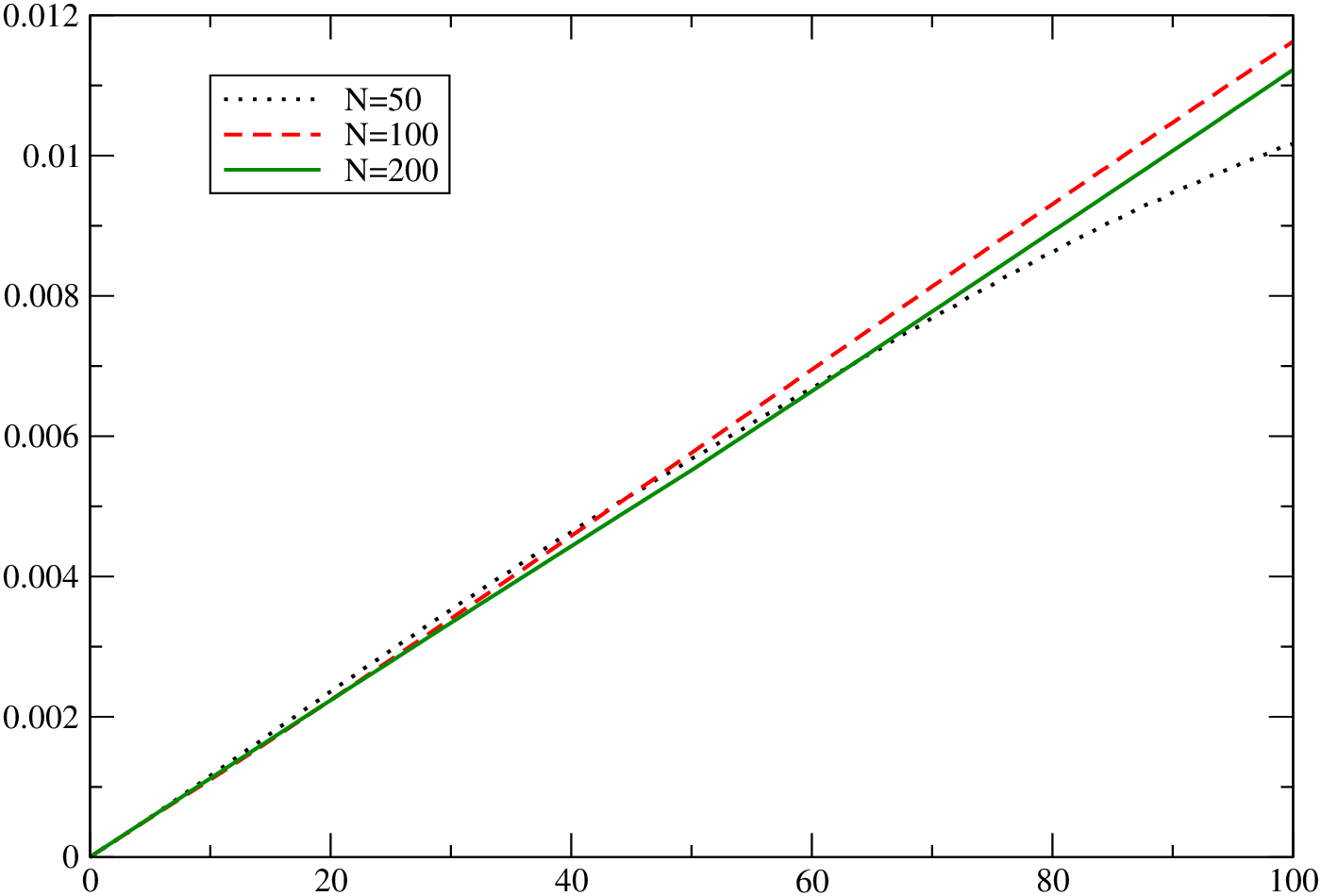}
  \caption{Plot of the rescaled error ${\cal E}_\rho(t)$ in $g_{xx}$ for 
           the 2D shifted gauge wave, with adjustment $b=1$.}
  \label{fig:GaugeWave2Dsh_new}
\end{figure}

\begin{table}[htbp]
  \begin{center}
    \begin{tabular}[c]{|c|c|}
\hline
   Test &
   Performance
    \\ \hline \hline
 $W$ versus $\hat W$ algorithm, 1D & $\hat W$ has smaller error  (Fig.~\ref{fig:WhWGaugeWave1Dsh})
\\ \hline
Stability of bare $\hat W$ algorithm, 1D & Conservation laws do not suppress exponential mode (Fig.~\ref{fig:hWGaugeWave1Dsh}, ~\ref{fig:GaugeWave1Dsh_gtt})
\\ \hline
 $\hat W$ versus $\hat V$ algorithm, 1D & In subluminal case, $\hat W$ has
much smaller error (Fig.~\ref{fig:WVGaugeWave1Dsh})
\\ \hline
 Adjusted $\hat W$ algorithm, with $c=1$ (eq.~\ref{eq:beladj}), 1D & 
 Adjustment suppresses the instability (Fig.~\ref{fig:CGaugeWave1Dsh}) 
\\ \hline
 $\hat W$ with $\lambda=1$ constraint damping (\ref{eq:constdamp}) and dissipation, 1D 
 & Ineffective, triggers large oscillations (Fig.~\ref{fig:DDGaugeWave1Dsh})
\\ \hline
 $\hat W$ adjusted with $c=1$ (\ref{eq:beladj}) or 
 $b = 1$ (\ref{eq:mmmmbeladj}), 1D  
 & Both adjustments suppress the instability  (Fig.~\ref{fig:GaugeWave1Dsh_new})
\\ \hline
 $\hat W$ algorithm adjusted with $b = 1$ (\ref{eq:mmmmbeladj}), convergence, 2D & 
 Long term second order convergence (Fig.~\ref{fig:GaugeWave2Dsh_new})
\\ \hline
\end{tabular}
    \caption{ Summary of tests performed 
              for the gauge wave with shift, with amplitude $A=0.5$. 
            }
    \label{tab:GaugeWavesh}
  \end{center}
\end{table}

\section{Conclusion}
\label{sec:concl}

Computational fluid dynamics developed into a mature field only after a long
struggle during which progress of lasting value emerged from the discriminating
use of model problems and standardized tests. An important potential payoff of
numerical relativity is the simulation of gravitational wave sources but sound
methodology and testing are essential in developing trustworthy codes. The
gauge wave tests presented here provide strong caution that a stable convergent
code is not sufficient to obtain long term simulations, as exhibited by the
problems introduced by long wavelength unstable modes in Figs.
\ref{fig:GaugeWave1D_gtt} and \ref{fig:GaugeWave1Dsh_gtt}. In the absence of
analytic solutions this caution extends to the simulation of binary black
holes. It would be of value to compare the results reported here for the
shifted gauge wave test with the performance of codes based upon different
hyperbolic reductions of the Einstein equations and different finite
difference approximations.

Our results show that discrete conservation laws for the principle part of the
evolution system are a good starting point for designing an algorithm but they
do not necessarily control nonlinear instabilities in the analytic problem. The
same holds true for the standard harmonic reduction of the Einstein tensor.  
In the case of periodic boundary conditions, the techniques introduced in this
paper are very effective in suppressing the long wavelength instabilities which
exist in the shifted gauge wave problem. We have extended these studies of the
shifted gauge wave to the initial-boundary value problem and a report is in
preparation~\cite{harm}.

\begin{acknowledgments}

We thank H-O. Kreiss for many valuable discussions and B. Schmidt for
improvements to the manuscript. This work was supported by the National Science
Foundation under grant PH-0244673 to the University of Pittsburgh. We used
computer time supplied by the Pittsburgh Supercomputing Center and we have
benefited from the use of the Cactus Computational Toolkit
(http://www.cactuscode.org).

\end{acknowledgments}


\begin{thebibliography}{40}


\bibitem{sci} R.~A. Matzner, H.~E. Seidel, S. L. Shapiro, L. Smarr, W-M
Suen, S.~A. Teukolsky, and J. Winicour, ``Geometry of a black hole
collision'', {\it Science} {\bf 270}, 941 (1995).

\bibitem{gce} A. Abrahams, L. Rezzolla, M. Rupright, et al,
``Schwarzschild perturbative gravitational wave extraction and outer
boundary conditions'', {\it Phys. Rev. Letters} {\bf 80}, 1812 (1998).

\bibitem{mbh} R. G\'{o}mez, L. Lehner, R.  Marsa and J. Winicour et al,
``Stable characteristic evolution of generic 3-dimensional
single-black-hole spacetimes'' {\it Phys. Rev. Letters} {\bf 80}, 3915
(1998).

\bibitem{kendpost} D.~E. Post and R.~P. Kendall,
{\it Internat. J. High Perf. Comput. Appl.}, {\bf 18}(4), 399 (2004).

\bibitem{postvot} D.~E. Post and L.~G. Votta, ``Computational science demands a
new paradigm'', {\em Physics Today}, {\bf 58}(1), 35 (2005).

\bibitem{mex1} The AppleswithApples Alliance, M. Alcubierre et al,
``Towards standard testbeds for numerical relativity'', 
{\it Class. Quantum Grav.}, {\bf 589} (2004).

\bibitem{awa} www.appleswithapples.org

\bibitem{mex2} The AppleswithApples Alliance,
``Implementation of standard testbeds for numerical relativity'',
in preparation.

\bibitem{friedrend} H. Friedrich and A.~D. Rendall, 
``The Cauchy problem for the Einstein equations'', in {\it Einstein's Field
Equations and Their Physical Implications: Selected Essays in Honour of 
J\"{u}rgen Ehlers}, ed. B.~G. Schmidt, Springer, Berlin (2000).

\bibitem{dedonder} T. de Donder, {\it La Gravifique Einsteinienne},
(Gauthiers-Villars, Paris, 1921).

\bibitem{Fock}
V. Fock,{\it The Theory of Space, Time and Gravitation},
(MacMillan, New York, 1964).

\bibitem{Choquet} Y. Foures-Bruhat,
``Theoreme d'existence pour certain systemes d'equations aux
derive\'{e}s partielles nonlinaires'',
{\it Acta Math} {\bf 88}, 141 (1952).

\bibitem{bab} M.~C. Babiuc, B. Szil\'{a}gyi and J. Winicour,
``Some mathematical problems in numerical relativity'',
{\it Lect. Notes Phys.} (to appear) gr-qc/0404092.


\bibitem{harl} B. Szil\'{a}gyi and J. Winicour,
``Well-posed initial-boundary evolution in general relativity'', 
{\it Phys. Rev.} {\bf D68}, 041501 (2003).

\bibitem{excis} B. Szil\'{a}gyi, H-O. Kreiss and J. Winicour, 
``Modeling the black hole excision problem'',
{\it Phys. Rev.} {\bf D71}, 104035 (2005).

\bibitem{z4}  C. Bona, T. Ledvinka, C. Palenzuela and M. Z\'{a}cek,
``General-covariant evolution formalism for numerical relativity'',
{\it Phys. Rev.} {\bf D67}, 104005 (2003).

\bibitem{Friedrich} H. Friedrich, ``Hyperbolic reductions for Einstein's
equations'', {\it Class. Quant. Grav.}, {\bf 13}, 1451 (1996).

\bibitem{Wald} R.~M. Wald, ``General Relativity'',
University of Chicago Press (1984). 

\bibitem{fisher}
A. E. Fisher and J. E. Marsden, {\em Comm. Math. Phys.},
{\bf 28}, 1 (1972).

\bibitem{pret1} F. Pretorius,
``Numerical Relativity Using a Generalized Harmonic Decomposition'',
{\it Class.Quant.Grav.} {\bf 22}, 425 (2005).

\bibitem{pret2} F. Pretorius,
``Evolution of Binary Black Hole Spacetimes'', {\em Phys.Rev.Lett.},
{\bf 95}, 121101 (2005).

\bibitem{constrdamp} C. Gundlach, J.~M. Martin-Garcia, G. Calabrese and I.
Hinder, ``Constraint damping in the Z4 formulation and harmonic gauge'', {\it
Class.Quant.Grav.} {\bf 22}, 3767 (2005).

\bibitem{alcsch} M. Alcubierre and B. Schutz, {\em J. Comput. Phys.}, {\bf 112},
44 (1994).

\bibitem{calab} G. Calabrese, 
``Finite differencing second order systems describing black hole
spacetimes'', {\em Phys.Rev. D} {\bf 71}, 027501 (2005). 


\bibitem{calabgund} G. Calabrese and C. Gundlach, 
``Discrete boundary treatment for the shifted wave equation'',
gr-qc/0509119.

\bibitem{live} M. Tiglio, L. Lehner and D. Neilsen,
``3D simulations of Einstein's equations:
symmetric hyperbolicity, live gauges and dynamic control of the constraints''
{\it Phys. Rev} {\bf D70}, 104018 (2004). 

\bibitem{discrenerg} L. Lehner, D. Neilsen, O. Reula and M. Tiglio,
``The discrete energy method in numerical relativity: Towards long-term
stability''
{\it Class.Quant.Grav.} {\bf 21} 5819 (2004).

\bibitem{sbp} G. Calabrese, L. Lehner, O. Reula, O. Sarbach, M. Tiglio,
``Summation by parts and dissipation for domains with excised regions'',
{\it Class.Quant.Grav.} {\bf 21}, 5735 (2004).

\bibitem{pitt} http://artemis.phyast.pitt.edu

\bibitem{harm} M.~C. Babiuc, B. Szil\'{a}gyi and J. Winicour,
``Harmonic initial-boundary evolution in general relativity''
(in preparation).

\end{thebibliography}
\end{document}